\documentclass[twocolumn,english,superscriptaddress,citeautoscript,showpacs,preprintnumbers,amsmath,amssymb,prd,floatfix,nofootinbib]{revtex4-1}
\pdfoutput=1
\usepackage{amsmath, amsthm, amssymb, graphicx,bm,amstext,subfigure, mathdots}
\usepackage{hyperref}
\usepackage{color}

\begin{document}
\title{Extra Attraction Generated by Spacetime Fluctuations}
\author{Qingdi Wang}
\affiliation{School of Physics, 
Xi'an Jiaotong University, Xi'an, Shaanxi 710049, China}
\begin{abstract}
We show that, due to the nonlinear nature of gravity, fluctuations in spacetime curvature generate additional gravitational attraction. This fluctuation-induced extra attraction was overlooked in the conventional understanding of the cosmological constant problem. If the quantum vacuum of matter fields possesses positive energy and negative pressure, it would produce enormous gravitational repulsion, resulting in a catastrophic explosion of the universe --- the acceleration of the universe's expansion would exceed the observed value by some 120 orders of magnitude. We argue that such an enormous repulsion produced by the violent matter fields vacuum can be completely suppressed by the even more substantial attraction generated by the zero-point fluctuations in the spacetime curvature. As a result, the predicted catastrophic explosion of the universe is averted.
\end{abstract}
\maketitle

\section{Introduction}
A fundamental aspect of general relativity is the highly nonlinear nature of the Einstein field equations, which govern the dynamical evolution of spacetime. This nonlinearity allows not only matter fields to act as a source of gravity, but also enables gravity to effectively source itself. Due to this mechanism, the nonlinear dynamics of curved spacetime, even in the absence of matter, can effectively generate additional gravitational forces. In short, gravity can create more gravity \cite{Markus}. A prime example of this phenomenon is the gravitational wave, a small disturbance in spacetime geometry that behaves similarly to electromagnetic radiation, carrying energy and momentum. Apart from the usual minor oscillations in spacetime geometry induced by the gravitational wave, its effective energy and momentum further curves the spacetime through which it propagates, generating slight additional net gravitational attraction (see, e.g., Misner, Thorne and Wheeler 1973, \cite{Misner:1973prb} p.961).

In this gravitational wave example, the additional net gravitational attraction caused by the spacetime disturbance is a small second-order effect and is usually ignored. However, one can reasonably expect that if the spacetime experiences large fluctuations, the nonlinear nature of gravity would result in significant attraction, an effect that can no longer be ignored. In fact, Khan and Penrose have shown that when two gravitational impulsive plane waves approaching each other from different directions, the additional attraction is so potent that it induces a significant focusing effect of each wave on the other, which eventually results in a singularity in spacetime \cite{Khan:1971vh}. In the first part of this paper (Sec. \ref{fluctuaion iuduced attraction}), we demonstrate that for generic fluctuations in spacetime curvature (not limited to small perturbations), the nonlinear nature of gravity consistently produce extra gravitational attraction.

Meanwhile, it is well established that all known physical fields must be described on a fundamental level by the principles of quantum theory (see, e.g., Wald  1984, \cite{Wald:1984rg} p.378). If one is to believe that the gravitational field adheres to the principles of quantum mechanics at the fundamental level, then the uncertainty principle implies that quantum fluctuations in spacetime curvature are inevitable. These fluctuations coexist with and are superposed on the large-scale, slowly varying curvature predicted by classical deterministic general relativity (see, e.g., Misner, Thorne and Wheeler 1973, \cite{Misner:1973prb} p.1192). In principle, even in the ground state of gravity (the gravitational field vacuum), spacetime curvature undergoes nonzero zero-point fluctuations, akin to how quantum matter fields experience zero-point fluctuations in the ground state of matter (the matter fields vacuum).

These quantum fluctuations in the spacetime curvature naturally give rise to the aforementioned gravitational attraction. Since quantum fluctuations intensify as one goes to smaller scales, we expect the small-scale quantum zero-point fluctuations in spacetime curvature to generate strong gravitational attraction. In the second part of this paper (Sec. \ref{quantum fluctuations} and \ref{extra attraction generated by quantum fluctuations}), we investigate the properties of the quantum zero-point fluctuations in spacetime curvature and demonstrate that quantum fluctuations indeed create strong gravitational attraction at small scales.

More importantly, these quantum fluctuation-induced extra attractions are not just interesting on their own. Their true importance lies in their critical role in understanding the cosmological constant problem. In the third part of this paper (Sec. \ref{ccp}), we highlight that the conventional formulation of the cosmological constant problem overlooks the extra gravitational attraction generated by the quantum fluctuations in the spacetime curvature and demonstrate that it plays a crucial role in addressing this problem.

More specifically, the cosmological constant problem essentially concerns the gravitational consequences of the enormous zero-point energy of the matter fields vacuum. Basically, the zero-point fluctuations in the matter fields vacuum contain tremendous zero-point energy, with its density on the order of Planck energy density if we trust the effective field theory up to the Planck energy scale. The equivalence principle of general relativity stipulates that every form of energy gravitates. On the one hand, it is widely assumed that the zero-point energy of the quantum matter fields vacuum gravitates as a cosmological constant. On the other hand, the discovery of the accelerating expansion of the universe in 1998 suggests that the Universe is driven by a positive cosmological constant, or equivalently, by the so-called dark energy with positive energy and negative pressure, which produces gravitational repulsion capable of causing the expansion of the universe to accelerate. For this reason, it has been naturally hoped that the zero-point energy of the quantum matter fields vacuum may act as this positive cosmological constant responsible for the acceleration of the universe's expansion. Unfortunately, the value of the zero-point energy predicted by quantum field theory is too large, and the gravitational repulsion it generates is expected to cause the expansion of the Universe to accelerate at a rate that would double its size on the Planck time scale ($\sim 10^{-43} s$).

This prediction is clearly flawed, as the observed accelerating expansion of the universe occurs at a rate that doubles its size on the time scale of $10$ billion years ($\sim 10^{17}s$), which is slower than the theoretical prediction by $60$ orders of magnitude. Correspondingly, the predicted value of the zero-point energy density from effective field theory calculation is larger than the observation by $120$ orders of magnitude. This is undoubtedly the largest discrepancy in all of science and has been described as the ``worst theoretical prediction in the history of physics" \cite{hobson2006general}. 

This discrepancy results in the long standing cosmological constant problem. In fact, the problem was already present since the birth of quantum field theory in 1920s \cite{Martin:2012bt}, much earlier than the discovery of the accelerating expansion of the universe in 1998. It lies at the cross road between quantum mechanics and general relativity and is widely regarded as one of the major obstacles to further progress in fundamental physics \cite{witten}.

In the literature, most proposed solutions to the cosmological constant problem attempt to either modify quantum field theory in some way to reduce the zero-point energy of the quantum matter fields or modify general relativity in some way to prevent the enormous energy from gravitating \cite{Rugh:2000ji, listofsolutions}. In contrast, we propose that the zero-point energy may indeed be as large as predicted by quantum field theory, and this immense energy does gravitate as mandated by general relativity. Nonetheless, the gravitation of this immense zero-point energy does not result in the disastrous explosion of the Universe. This is because the prediction of catastrophic explosion considered only the large repulsion generated by the zero-point energy of quantum matter fields vacuum, overlooking the large attraction produced by the zero-point fluctuations in spacetime curvature. Our analysis indicates that the attraction generated by the zero-point fluctuations in spacetime curvature is greater than the repulsion arising from the zero-point energy of the matter fields vacuum, thereby preventing the anticipated catastrophic explosion.

Previously, we have also tried to tackle the cosmological constant problem in a series of papers \cite{PhysRevD.95.103504}, \cite{PhysRevD.98.063506}, \cite{PhysRevD.102.023537} and \cite{PhysRevLett.125.051301}. In these papers, we noticed that the magnitude of the zero-point fluctuation itself also fluctuates, leading to a constantly fluctuating and extremely inhomogeneous vacuum energy density of quantum matter fields. As a result, the small-scale structure of spacetime sourced by matter fields vacuum becomes fundamentally different. Rather than the space expanding or contracting as a whole, as described by the usual homogeneous FLRW metric, locally, at each point, the space oscillates alternatively between expansion and contraction, with varying phases of oscillation at neighboring points. These local expansions and contractions are violent at small scales. However, upon averaging over larger macroscopic scales, their effects nearly cancel out, except for a slight dominance of expansions over contractions, due to a subtle parametric resonance induced by zero-point fluctuations in the matter fields vacuum. This naturally leads to the slowly accelerating expansion of the universe on the cosmological scale.

However, in these papers, either the gravitational effects of the quantum matter fields were predetermined to be attractive \cite{PhysRevD.95.103504, PhysRevD.98.063506}, or an exceedingly large negative cosmological constant was introduced to provide the necessary attraction \cite{PhysRevD.102.023537, PhysRevLett.125.051301}. In this paper, we have attained similar small-scale spacetime structure as previously explored, but with a significant advance. The needed attraction emerges naturally from the quantum zero-point fluctuations of the spacetime curvature. In addition, our finding indicates that the slowly accelerating expansion of the universe is primarily driven by zero-point fluctuations in the gravitational vacuum, rather than the matter fields vacuum.

This paper is organized as follows: in Sec. \ref{fluctuaion iuduced attraction}, we demonstrate that fluctuations in the Riemann curvature always generate gravitational attractions; in Sec. \ref{quantum fluctuations}, we discuss the sources and the properties of the quantum fluctuations in the Riemann curvature; in Sec. \ref{extra attraction generated by quantum fluctuations}, we investigate the property of the quantum fluctuation-induced attractions; in Sec. \ref{ccp}, we apply the results we obtained in the previous sections to the cosmological constant problem; in Sec. \ref{small scale structure}, we discuss some notable issues of the cosmological constant problem and the differences between the work presented in this paper and the previous ones; in Sec. \ref{summary}, we summarize our results.

Throughout the paper, we use the units $\hbar=c=1$ and the metric signature $(-, +, +, +)$.

\section{The Fluctuation Induced Extra Attraction}\label{fluctuaion iuduced attraction}
Consider a spacetime exhibiting fluctuations, where the spacetime metric can be decomposed into two components: a slowly varying average term, referred to as the ``background" term, and a rapidly varying fluctuation term. This decomposition is represented by the equation:
\begin{equation}\label{metric split}
g_{ab}=g_{ab}^{(\mathrm{B})}+g_{ab}^{(\mathrm{F})},
\end{equation}
where the fluctuation term $g_{ab}^{(\mathrm{F})}$ has a zero average, representing the deviations from the background metric $g_{ab}^{(\mathrm{B})}$. It should be noted that the magnitude of the fluctuation term $g_{ab}^{(\mathrm{F})}$ is not necessarily assumed to be small.

One of the essential features of gravity is that the Einstein equations, unlike for example Maxwell's equations, are highly nonlinear. If we assume that Einstein equations hold exactly for $g_{ab}$
\begin{equation}
G_{ab}\left[g^{(\mathrm{B})}+g^{(\mathrm{F})}\right]=8\pi G T_{ab},
\end{equation}
we can then, in principle, derive an effective Einstein equations holding for the background metric $g_{ab}^{(\mathrm{B})}$ by averaging out the effect of the rapidly varying fluctuations $g_{ab}^{(\mathrm{F})}$ \cite{Bolejko:2016qku, Ellis_2011}:
\begin{equation}\label{effective Einstein equations}
G_{ab}\left[g^{(\mathrm{B})}\right]=8\pi G \left(\left\langle T_{ab}\right\rangle+t_{ab}\left[g^{(\mathrm{F})}\right]\right).
\end{equation}
The back reaction term $t_{ab}$ in \eqref{effective Einstein equations} emerges due to the nonlinear structure of Einstein equations. It is generated by the fluctuations $g_{ab}^{(\mathrm{F})}$ and acts as a stress energy tensor\footnote{Strictly speaking, the back reaction $t_{ab}$ may not precisely behave as a stress energy tensor, since in the general non-perturbative scenario, it may not satisfy the conservation equations $\nabla^at_{ab}=0$.} which effectively creates the background curvature.

A well-known example is the gravitational wave \cite{PhysRev.166.1272}, where the fluctuation term $g_{ab}^{(\mathrm{F})}$ in \eqref{metric split} serves as a small perturbation. In the transverse-traceless gauge, the effective stress-energy tensor of the gravitational wave takes the form
\begin{equation}
t_{\mu\nu}^{(\mathrm{GW})}=\frac{1}{32\pi G}\left\langle\nabla_{\mu}g_{\alpha\beta}^{(\mathrm{F})}\nabla_{\nu}g^{(\mathrm{F})\alpha\beta}\right\rangle, \quad\text{(TT gauge)},
\end{equation}
where the angle brackets $\langle\rangle$ mean averaging over several wavelengths, the derivative $\nabla$ is associated with the background metric $g^{\mathrm{(B)}}_{ab}$ (see, e.g., p.970 of \cite{Misner:1973prb}). For a general plane wave propagating in the $z$-direction, we have the metric perturbations (see, e.g., p.1004-1005 of \cite{Misner:1973prb}):
\begin{eqnarray}
g^{(\mathrm{F})}_{xx}&=&-g^{(\mathrm{F})}_{yy}=A_{+}(t-z),\label{plane wave metric 1}\\
g^{(\mathrm{F})}_{xy}&=&g^{(\mathrm{F})}_{yx}=A_{\times}(t-z),\label{plane wave meric 2}
\end{eqnarray}
and the effective stress energy
\begin{equation}\label{stress energy of gravitational wave}
t_{00}^{(\mathrm{GW})}=t_{zz}^{(\mathrm{GW})}=-t_{0z}^{(\mathrm{GW})}=\frac{1}{16\pi G}\left\langle \dot{A}^2_{+}+\dot{A}^2_{\times}\right\rangle,
\end{equation}
where $A_{+}, A_{\times}\ll 1$ are two general oscillating functions with the properties
\begin{eqnarray}
&&\Big\langle A_{+}\Big\rangle=\left\langle \dot{A}_{+}\right\rangle=\left\langle \ddot{A}_{+}\right\rangle=0, \label{property 1}\\
&&\Big\langle A_{\times}\Big\rangle=\left\langle \dot{A}_{\times}\right\rangle=\left\langle \ddot{A}_{\times}\right\rangle=0. \label{property 2}
\end{eqnarray}

In the case of gravitational waves, the matter fields contribution $T_{ab}$ in the effective Einstein equations \eqref{effective Einstein equations} is absent and we have
\begin{equation}\label{effective Einstein equations for gravity wave}
G^{(\mathrm{GW})}_{ab}\left[g^{(\mathrm{B})}\right]=8\pi G t^{(\mathrm{GW})}_{ab}\left[g^{(\mathrm{F})}\right].
\end{equation}

It is important to note that the expression \eqref{stress energy of gravitational wave} for the effective stress energy $t_{ab}$ of the plane gravitational wave has positive energy and pressure. Since positive energy and pressure produces gravitational attractions, the gravitational wave fluctuations $g^{(\mathrm{F})}$ would curve the background $g^{(\mathrm{B})}$ through the effective Einstein equations \eqref{effective Einstein equations for gravity wave} in a way that generates attractions. When a gravitational wave passes through a set of free falling test particles, in addition to the ordinary oscillatory motions, in the mean time the gravitational wave would generate an extra net overall attraction which tends to pull these test particles closer.

In this gravitational wave example, the fluctuations $g^{(\mathrm{F})}$ are small. The additional net attraction resulting from these fluctuations is a second-order effect and is typically disregarded. This raises a natural question: for general fluctuations $g^{\mathrm{(F)}}$ that are not necessarily small perturbations, does the back reaction $t_{ab}$ on the right hand side of \eqref{effective Einstein equations} still give rise to gravitational attractions? If so, what is the strength of this attraction? Further more, can it still be neglected when the fluctuations are strong?

One might hope that these questions can be addressed by directly analyzing the structure of $t_{ab}$, similar to the above gravitational wave example. However, due to the complexity of the Einstein equations, it is even technically difficult to obtain an explicit expression for $t_{ab}$ when the fluctuation is not a perturbation.

To overcome this challenge, we approach these questions from a different angle. Instead of directly examining the structure of $t_{ab}$, we investigate the relative motion between infinitesimally nearby free falling test particles (i.e., neighboring geodesics) moving in the fluctuating spacetime. By analyzing the geodesic deviation equation, we demonstrate that in the next subsection that the fluctuations always generate extra tidal attractions between these nearby free falling test particles (neighboring geodesics). This finding implies that the back reaction $t_{ab}$ in the effective Einstein equations \eqref{effective Einstein equations} consistently causes the background $g_{ab}^{(\mathrm{B})}$ to curve in a manner that generates gravitational attractions. Notably, these extra attractions become substantial when the fluctuations are significant and thus should not be overlooked.

\subsection{The Extra Tidal Attraction}\label{attraction}
For convenience, in the following we will employ the notation $\overline{A}$ and $\xi_A$ to denote the average and fluctuation of a given physical quantity $A$, respectively.

By definition, the spacetime curvature also undergoes fluctuations alongside the metric fluctuations. Similar to the metric split represented by \eqref{metric split}, we assume that the Riemann curvature can be separated into a slowly varying average ``background" term plus a rapidly fluctuating term\footnote{Note that due to the nonlinearity of curvature, the average curvature $\overline{R}^a_{bcd}$ in \eqref{Riemann curvature fluctuation} is not equal to the curvature associated with the background metric $g_{ab}^{(\mathrm{B})}$ in \eqref{metric split}, i.e., $\overline{R}^a_{bcd}\neq R^a_{bcd}\left[g^{(\mathrm{B})}\right]$.}
\begin{equation}\label{Riemann curvature fluctuation}
R^a_{bcd}=\overline{R}^a_{bcd}+\xi_{R^a_{bcd}},
\end{equation}
where $\xi_{R^a_{bcd}}$ has zero average and represents the fluctuations about the background curvature $\overline{R}^a_{bcd}$.

The Riemann curvature relates to the tidal accelerations between neighboring geodesics through the geodesic deviation equation (see, e.g., pages 46-47 of Wald 1984, \cite{Wald:1984rg})
\begin{equation}\label{original geodesic deviation equation}
a^a=T^c\nabla_c\left(T^b\nabla_bX^a\right)=-R^a_{cbd}X^bT^cT^d.
\end{equation}
Here we use the same notation as in \cite{Wald:1984rg} that we consider a one parameter family of geodesics $\gamma_s$ parameterized by affine parameter $t$; $T^a=(\partial/\partial t)^a$ in \eqref{original geodesic deviation equation} is the tangent to $\gamma_s$, $X^a=(\partial/\partial s)^a$ in \eqref{original geodesic deviation equation} is the deviation vector which represents the relative displacement from $\gamma_s$ to an infinitesimally nearby geodesic,  $a^a$ in \eqref{original geodesic deviation equation} is the tidal acceleration of an infinitesimally nearby geodesic relative to $\gamma_s$ (as in \cite{Wald:1984rg}, we have chosen proper parameterization such that $X^aT_a=0$ everywhere).

By the geodesic deviation equation \eqref{original geodesic deviation equation}, the presence of the Riemann curvature fluctuations $\xi_{R^a_{bcd}}$ in \eqref{Riemann curvature fluctuation} cause fluctuations in the tidal forces and in the relative accelerations between neighboring geodesics. These fluctuating tidal forces jiggle the trajectory of the test particle. As a result, $X^a$ will traverse a smooth path and at the same time execute oscillations about that path. Accordingly, we express $X^a$ as a sum:
\begin{equation}\label{xa decomposition}
X^a=\overline{X}^a+\xi_{X^a},
\end{equation}
where $\overline{X}^a$ represents the average smooth path, $\xi_{X^a}$ represents the oscillations.

One might think at this point that $\overline{X}^a$ is just the path that $X^a$ would execute in the ``background" spacetime with curvature $\overline{R}^a_{bcd}$ alone (i.e., if $\xi_{R^a_{bcd}}=0$), and the curvature fluctuations $\xi_{R^a_{bcd}}$ just jiggle $X^a$ about that path.

Surprisingly, $\overline{X}^a$ would execute a different path. Comparing to the path that $X^a$ would execute in the ``background" spacetime with the curvature $\overline{R}^a_{bcd}$ alone, $\overline{X}^a$ would experience extra accelerations towards $\gamma_s$. In other words, the net effect of the Riemann curvature fluctuations $\xi_{R^a_{bcd}}$ on the average path $\overline{X}^a$ is not zero. It would generate extra tidal attractions which tend to pull the geodesics closer.

\subsubsection{The Geodesic Deviation Equation in Local Inertial Frame}\label{Geodesic deviation equation}
Without loss of generality, we pick up an arbitrary geodesic from the one parameter family $\gamma_s$ and reparameterize it by its proper time $\tau$. For convenience, we denote this geodesic simply by $\gamma$, its tangent $T^a$ by $(e_0)^a$, which is normalized to unit length so that $g_{ab}(e_0)^a(e_0)^b=-1$. Then the geodesic deviation equation \eqref{original geodesic deviation equation} can be expressed as:
\begin{equation}\label{geodesic deviation equation}
(e_0)^c\nabla_c\left((e_0)^b\nabla_bX^a\right)=-R^a_{cbd}(e_0)^c(e_0)^dX^b.
\end{equation}
A solution $X^a$ of \eqref{geodesic deviation equation} is called a Jacobi field on $\gamma$. It describes how the deviation vector $X^a$ evolves along $\gamma$.

In order to investigate the property of the Jacobi field $X^a$ along $\gamma$, we introduce an orthonormal basis of spatial vectors $(e_1)^a, (e_2)^a, (e_3)^a$ orthogonal to $(e_0)^a$ and parallelly propagated along $\gamma$. The frame $\{(e_0)^a, (e_1)^a, (e_2)^a, (e_3)^a\}$ constructed in this way is orthonormal everywhere along $\gamma$, i.e., we have
\begin{equation}
g_{ab}(e_\mu)^a(e_{\nu})^a=\eta_{\mu\nu},
\end{equation}
with $\eta_{\mu\nu}=\mathrm{diag}\left(-1, 1, 1, 1\right)$. This frame is just $\gamma$'s local inertial frame.

We express $X^a$ in terms of the orthonormal basis $\{(e_1)^a, (e_2)^a, (e_3)^a\}$ as:
\begin{equation}\label{xa expansion}
X^a=X^1(e_1)^a+X^2(e_2)^a+X^3(e_3)^a.
\end{equation}
By substituting the above expression \eqref{xa expansion} into the geodesic deviation equation \eqref{geodesic deviation equation}, we obtain the ordinary differential equations for the components $X^i$ of the deviation vector $X^a$:
\begin{equation}\label{component form of geodesic deviation equation}
\frac{d^2X^i}{d\tau^2}=-\sum_{j=1}^3  R^i_{0j0}X^j, \quad i=1, 2, 3,
\end{equation}
where the components $R^i_{0j0}$ of the Riemann curvature tensor $R^a_{bcd}$ are defined as:
\begin{equation}
R^i_{0j0}=R^a_{cbd}(e_0)^c(e_0)^d(e^i)_a(e_j)^b.
\end{equation}

For convenience, we can express \eqref{component form of geodesic deviation equation} in a matrix form as:
\begin{equation}\label{matrix form}
\frac{d^2\mathbf{X}}{d\tau^2}=-\mathrm{R}\mathbf{X},
\end{equation}
where
\begin{equation}
\mathbf{X}=\begin{pmatrix}
X^1\\
X^2\\
X^3
\end{pmatrix},\quad   
\mathrm{R}=\begin{pmatrix}
R^1_{010} & R^1_{020} & R^1_{030} \\
R^2_{010} & R^2_{020} & R^2_{030} \\
R^3_{010} & R^3_{020} & R^3_{030}
\end{pmatrix}.\label{matrix definition}
\end{equation}
In the local inertial frame $\{(e_0)^a, (e_1)^a, (e_2)^a, (e_3)^a\}$, the Riemann curvature components satisfy
\begin{equation}
R^i_{0j0}=R_{0i0j}=R_{0j0i}=R^j_{0i0},
\end{equation}
which implies that the curvature matrix $\mathrm{R}$ in \eqref{matrix form} is symmetric.

Along the trajectory of $\gamma$, the curvature component $R^i_{0j0}$ in \eqref{component form of geodesic deviation equation} or the matrix $\mathrm{R}$ in \eqref{matrix form} varies with the proper time $\tau$. Thus, equation \eqref{component form of geodesic deviation equation} or \eqref{matrix form} describes the motion of an infinitesimally nearby free falling test particle in the local inertial frame of $\gamma$, subject to a time-dependent potential:
\begin{equation}
\frac{d^2\mathbf{X}}{d\tau^2}=-\nabla U,
\end{equation}
where
\begin{equation}\label{potential definition}
U(\tau, \mathbf{X})=\frac{1}{2}\sum_{i=1}^3X^iR^i_{0j0}(\tau)X^j=\frac{1}{2}\mathbf{X}^t\mathrm{R}(\tau)\mathbf{X}.
\end{equation}
Here, the gradient $\nabla=(\frac{\partial}{\partial X^1}, \frac{\partial}{\partial X^2}, \frac{\partial}{\partial X^3})$ and $\mathbf{X}^t$ denotes the transpose of $\mathbf{X}$.

\subsubsection{The Ponderomotive-like Extra Attraction Force}
Suppose that $\xi_{R_{abcd}}$ in \eqref{Riemann curvature fluctuation} is rapidly fluctuating around zero in the sense that, along the timelike geodesic $\gamma$, the components $\xi_{R^{\mu}_{\nu\lambda\rho}}$ of $\xi_{R^a_{bcd}}$ in $\gamma$'s local inertial frame $\{(e_0)^a, (e_1)^a, (e_2)^a, (e_3)^a\}$ vary with $\gamma$'s proper time $\tau$ on a time scale $T$ much smaller than the time scale of variation of the ``background" spacetime metric, which is characterized by the typical value of $1/|\overline{R}^{\mu}_{\nu\lambda\rho}|^{1/2}$,
\begin{equation}\label{time scale of variation}
T\ll \text{typical value of }1/|\overline{R}^{\mu}_{\nu\lambda\rho}|^{1/2},
\end{equation}
where $\overline{R}^{\mu}_{\nu\lambda\rho}$ are the components of $\overline{R}^a_{bcd}$ in the frame $\{(e_0)^a, (e_1)^a, (e_2)^a, (e_3)^a\}$.

Restricting the Riemann curvature split \eqref{Riemann curvature fluctuation} on the trajectory of $\gamma$, we can express the $3\times 3$ curvature matrix $\mathrm{R}$ defined by \eqref{matrix definition} along $\gamma$ as
\begin{equation}
\mathrm{R}(\tau)=\overline{\mathrm{R}}(\tau)+\xi_{\mathrm{R}}(\tau),\label{2}
\end{equation}
where $\overline{\mathrm{R}}(\tau)$ represents the slowly varying average coming from the ``background" curvature term $\overline{R}^a_{bcd}$ and $\xi_{\mathrm{R}}(\tau)$ represents the rapid variation coming from the fluctuation term $\xi_{R^a_{bcd}}$.

Accordingly, the potential $U$ defined by \eqref{potential definition} can be decomposed into two components: a slowly varying background potential and a rapidly fluctuating potential, given by
\begin{equation}
U=\overline{U}+\xi_U,
\end{equation}
where
\begin{equation}\label{Ubar definition}
\overline{U}=\frac{1}{2}\mathbf{X}^t\overline{\mathrm{R}}(\tau)\mathbf{X}, \quad \xi_U=\frac{1}{2}\mathbf{X}^t\xi_{\mathrm{R}}(\tau)\mathbf{X}.
\end{equation}

The decomposition \eqref{xa decomposition} of the deviation vector $X^a$ can be expressed in the local inertial frame of $\gamma$ as
\begin{equation}
\mathbf{X}(\tau)=\overline{\mathbf{X}}(\tau)+\mathbf{\xi}_{\mathbf{X}}(\tau),\label{1}
\end{equation}
where $\overline{\mathbf{X}}(\tau)$ represents the average smooth path and $\mathbf{\xi}_{\mathbf{X}}(\tau)$ represents the oscillations about that path.

To be more precise, we define $\overline{\mathrm{R}}(\tau)$ and $\overline{\mathbf{X}}(\tau)$ as the averages of $\mathrm{R}$ and $\mathbf{X}$ over the time interval $(\tau-T/2, \tau+T/2)$:
\begin{eqnarray}
\overline{\mathrm{R}}(\tau)&=&\frac{1}{T}\int_{\tau-\frac{T}{2}}^{\tau+\frac{T}{2}}\mathrm{R}(\tau')d\tau',\\
\overline{\mathbf{X}}(\tau)&=&\frac{1}{T}\int_{\tau-\frac{T}{2}}^{\tau+\frac{T}{2}}\mathbf{X}(\tau')d\tau'.\label{average definition}
\end{eqnarray}

We will derive an equation that governs the evolution of $\overline{\mathbf{X}}(\tau)$. As mentioned earlier in this section, it will not simply follow the motion subject to the potential $\overline{U}$ alone:
\begin{equation}
\frac{d^2\overline{\mathbf{X}}}{d\tau^2}\neq -\nabla\overline{U}=-\overline{\mathrm{R}}\,\overline{\mathbf{X}}.
\end{equation}
Instead, the fluctuation $\xi_{\mathrm{R}}$ would generate an extra effective attractive potential which tends to decrease $\overline{\mathbf{X}}$. 

Physically, this phenomenon occurs because the fluctuating tidal force $-\mathrm{R}\mathbf{X}$ in the geodesic deviation equation \eqref{matrix form} is position dependent. As a result, equation \eqref{matrix form} describes the motion of a particle in an inhomogeneously oscillating field. It is well established that a particle moving in an inhomogeneously oscillating field experiences a ponderomotive-like force, which causes the particle to move towards the area of the weaker field strength, rather than oscillating around an initial point as happens in a homogeneous field \cite{MichaelFowler, Lundin2006}.

More precisely, the particle exhibits fast oscillations in response to the rapid fluctuation term $\xi_{\mathrm{R}}$. It sees a greater magnitude of restoring force during the half of the oscillation period while it is in the area of larger $|\mathbf{X}|$. The restoring force during the other half of the oscillation period while it is in the area of smaller $|\mathbf{X}|$ is smaller. So over a complete cycle, there is a net force which tends to make the particle move towards the area of smaller $|\mathbf{X}|$.

Next we derive a formula for the magnitude of this extra attraction force. This derivation follows the same principle as in Landau and Lifshitz 1976, \cite{LANDAU197658} p.93-94. The physical mechanism behind the Ponderomotive-like force is partially inspired by Smolyaninov's work \cite{Smolyaninov:2019tiz} and Lozano and Mazzitelli's work \cite{Lozano:2020xga}.

Substituting \eqref{2} and \eqref{1} into \eqref{matrix form} we have
\begin{equation}\label{slow fast expansion}
\frac{d^2\overline{\mathbf{X}}}{d\tau^2}+\frac{d^2\xi_{\mathbf{X}}}{d\tau^2}=-\overline{\mathrm{R}}\,\overline{\mathbf{X}}-\xi_{\mathrm{R}}\overline{\mathbf{X}}-\overline{\mathrm{R}}\xi_{\mathbf{X}}-\xi_{\mathrm{R}}\xi_{\mathbf{X}}.
\end{equation}

The magnitude of the curvature fluctuations $\xi_{\mathrm{R}}$ in the above equation \eqref{slow fast expansion} is not assumed to be small, but the amplitude of the geodesic oscillations $\xi_{\mathbf{X}}$ in \eqref{slow fast expansion} can be assumed to be small. That is because, by the assumption we made earlier in this subsection about the time scale of variation of $\xi_{R^{\mu}_{\nu\lambda\rho}}$, $\xi_{\mathrm{R}}$ fluctuates so fast that there is no enough time for $\xi_{\mathbf{X}}$ to grow significantly\footnote{This point might be easier to understand by the following simple illustrative example: the motion of a particle given by equation $\ddot{x}=-\kappa\cos\omega t$. The solution to this equation is $x=c_1+c_2t+\kappa\cos\omega t/\omega^2$, where $c_1$, $c_2$ are constants determined by initial conditions. The oscillatory part of motion of $x$ is given by $\xi_x=\kappa\cos\omega t/\omega^2$, which is inversely proportional to $\omega^2$. Thus higher frequency leads to smaller amplitude of oscillations.}.

The equation \eqref{slow fast expansion} involves both oscillatory and smooth terms, which must evidently be separately equal. For the oscillatory terms we can put simply
\begin{equation}\label{oscillating equation}
\frac{d^2\xi_{\mathbf{X}}}{d\tau^2}=-\xi_{\mathrm{R}}\overline{\mathbf{X}},
\end{equation}
since other terms contain the small factor $\xi_{\mathbf{X}}$ and are therefore of higher order of smallnesses. On the time scale $T$, $\overline{\mathbf{X}}(\tau)$ changes only slightly. Thus we can regard $\overline{\mathbf{X}}$ in the above equation \eqref{oscillating equation} as a constant. Integrating \eqref{oscillating equation} gives
\begin{equation}\label{small oscillation}
\xi_{\mathbf{X}}=-\mathrm{D}\overline{\mathbf{X}},
\end{equation}
where
\begin{equation}
\mathrm{D}(\tau)=\int^{\tau}\mathrm{V}d\tau',
\end{equation}
with
\begin{equation}\label{V definition}
\mathrm{V}(\tau)=\int^{\tau}\xi_{\mathrm{R}}d\tau'.
\end{equation}

Substituting \eqref{small oscillation} into \eqref{matrix form} (or \eqref{slow fast expansion}) gives
\begin{equation}\label{equaiton before average}
\frac{d^2\mathbf{X}}{d\tau^2}=-\mathrm{R}\overline{\mathbf{X}}-\mathrm{V}^2\overline{\mathbf{X}}+\overline{\mathrm{R}}\mathrm{D}\overline{\mathbf{X}}+\frac{d\left(\mathrm{V}\mathrm{D}\right)}{d\tau}\overline{\mathbf{X}},
\end{equation}
where we have used the relation
\begin{equation}
\xi_{\mathrm{R}}\mathrm{D}=\frac{d}{d\tau}\left(\mathrm{V}\mathrm{D}\right)-\mathrm{V}^2.
\end{equation}

Next we take average on both sides of the above equation \eqref{equaiton before average} over the time interval $(\tau-T/2, \tau+T/2)$ (in the sense defined by \eqref{average definition}). Since the mean values of $\mathrm{D}$ and $d(\mathrm{VD})/d\tau$ are zero, we obtain the equation for the evolution of $\overline{\mathbf{X}}$,
\begin{equation}\label{average evolution equation for X with fluctuations}
\frac{d^2\overline{\mathbf{X}}}{d\tau^2}=-\left(\overline{\mathrm{R}}+\overline{\mathrm{V}^2}\right)\overline{\mathbf{X}}\neq -\nabla\overline{U},
\end{equation}
where
\begin{equation}\label{Vsqrure average definition}
\overline{\mathrm{V}^2}(\tau)=\frac{1}{T}\int_{\tau-\frac{T}{2}}^{\tau+\frac{T}{2}}\mathrm{V}^2(\tau')d\tau'.
\end{equation}

If there are no curvature fluctuations, i.e., if $\xi_{\mathrm{R}}=0$, we have $\overline{\mathrm{R}}=\mathrm{R}$, $\overline{\mathbf{X}}=\mathbf{X}$ and thus
\begin{equation}\label{average evolution equation for X without fluctuations}
\frac{d^2\overline{\mathbf{X}}}{d\tau^2}=\frac{d^2\mathbf{X}}{d\tau^2}=-\mathrm{R}\mathbf{X}=-\overline{\mathrm{R}}\,\overline{\mathbf{X}}.
\end{equation}

When comparing the evolution equation \eqref{average evolution equation for X with fluctuations} in spacetime with curvature fluctuations and the evolution equation \eqref{average evolution equation for X without fluctuations} in spacetime without curvature fluctuations, the extra term
\begin{equation}\label{the extra attraction force}
\mathbf{A}_{\mathrm{extra}}=-\overline{\mathrm{V^2}}\,\overline{\mathbf{X}}
\end{equation}
represents the extra tidal attraction generated by the Riemann curvature fluctuations. This term can be expressed as
\begin{eqnarray}
\mathbf{A}_{\mathrm{extra}}=-\nabla U_{\mathrm{eff}},
\end{eqnarray}
where the gradient $\nabla=\left(\frac{\partial}{\partial\overline{X^1}},\frac{\partial}{\partial\overline{X^2}},\frac{\partial}{\partial\overline{X^3}}\right)$ and the effective potential $U_{\mathrm{eff}}$ is defined by
\begin{equation}
U_{\mathrm{eff}}\left(\overline{\mathbf{X}}\right)=\frac{1}{2}\mathbf{\overline{X}}^{\mathrm{t}}\,\overline{\mathrm{V^2}}\,\mathbf{\overline{X}}=\frac{1}{2}\overline{\left(\frac{d\xi_{\mathbf{X}}}{d\tau}\right)^2},
\end{equation}
representing the mean kinetic energy of the oscillatory motion $\xi_{\mathbf{X}}$ about the average path $\overline{\mathbf{X}}$.

The $(i,j)$ element of the matrix $\overline{\mathrm{V}^2}$ can be expressed as
\begin{eqnarray}\label{V^2 elements}
\left(\overline{\mathrm{V}^2}\right)_{ij}&=&\sum_{k=1}^3\overline{\mathrm{V}_{ik}\mathrm{V}_{kj}}\nonumber\\
&=&\begin{cases}
    \sum_{k=1}^3\overline{\mathrm{V}_{ik}^2},       & \quad \text{if } i=j, \\
    \sum_{k=1}^3\overline{\mathrm{V}_{ik}\mathrm{V}_{jk}}  & \quad \text{if } i\neq j.
  \end{cases}
\end{eqnarray}

If the fluctuations in $\mathrm{R}$ exhibit no preferred directions, i.e., if we have
\begin{eqnarray}
&&\overline{\xi_{\mathrm{R}_{11}}^2}=\overline{\xi_{\mathrm{R}_{22}}^2}=\overline{\xi_{\mathrm{R}_{33}}^2},\\
&&\overline{\xi_{\mathrm{R}_{12}}^2}=\overline{\xi_{\mathrm{R}_{13}}^2}=\overline{\xi_{\mathrm{R}_{23}}^2},
\end{eqnarray}
then according to \eqref{V definition}, the follow relations hold:
\begin{eqnarray}
\overline{\left(\mathrm{V}_{11}\right)^2}=\overline{\left(\mathrm{V}_{22}\right)^2}=\overline{\left(\mathrm{V}_{33}\right)^2},\label{Vij properties1}\\
\overline{\left(\mathrm{V}_{12}\right)^2}=\overline{\left(\mathrm{V}_{13}\right)^2}=\overline{\left(\mathrm{V}_{23}\right)^2}.\label{Vij properties2}
\end{eqnarray}

If there are no special relations between the fluctuations in $\mathrm{V}_{ik}$ and $\mathrm{V}_{jk}$, we have
\begin{equation}\label{Vikjk no relation}
\overline{\mathrm{V}_{ik}\mathrm{V}_{jk}}=0, \quad i\neq j.
\end{equation}
Then from \eqref{V^2 elements} we have
\begin{equation}
\left(\overline{\mathrm{V}^2}\right)_{ij}=v^2\delta_{ij},
\end{equation}
where
\begin{equation}\label{vsquare definition}
v^2=\sum_{k=1}^3\overline{\mathrm{V}_{ik}^2}.
\end{equation}
Or in matrix form,
\begin{equation}\label{V^2 matrix form}
\overline{\mathrm{V}^2}=v^2\mathrm{I},
\end{equation}
where $\mathrm{I}=\mathrm{diag}\left(1, 1, 1\right)$ represents the identity matrix.

Inserting \eqref{V^2 matrix form} into \eqref{the extra attraction force} we obtain the extra tidal attraction towards the fiducial geodesic $\gamma$:
\begin{equation}\label{Extra force}
\mathbf{A}_{\mathrm{extra}}=-v^2\overline{\mathbf{X}}.
\end{equation}

\subsubsection{The Gravitational Wave Example}
In order to gain a deeper understanding the physical mechanism involved, we apply the result we obtained in the last subsubsection to the fluctuations of the plane gravitational wave described by \eqref{plane wave metric 1} and \eqref{plane wave meric 2}, which propagates in the $z$-direction ($X^3$-direction).

In this example, the Riemann curvature takes the form:
\begin{eqnarray}
R^{1(\mathrm{GW})}_{010}&=&-R^{2(\mathrm{GW})}_{020}=-\frac{1}{2}\ddot{A}_{+},\\
R^{1(\mathrm{GW})}_{020}&=&R^{2(\mathrm{GW})}_{010}=-\frac{1}{2}\ddot{A}_{\times}.
\end{eqnarray}
Then from the properties \eqref{property 1} and \eqref{property 2}, we have $\overline{\mathrm{R}}=0$ and thus the curvature matrix
\begin{equation}\label{wave curvature matrix}
\mathrm{R}=\xi_{\mathrm{R}}=\begin{pmatrix}
-\frac{1}{2}\ddot{A}_{+} & -\frac{1}{2}\ddot{A}_{\times} & 0 \\
-\frac{1}{2}\ddot{A}_{\times} & +\frac{1}{2}\ddot{A}_{+} & 0 \\
0 & 0 & 0
\end{pmatrix}.
\end{equation}
Substituting \eqref{wave curvature matrix} into the definition \eqref{V definition} for $\mathrm{V}$ yields
\begin{equation}\label{V wave}
\mathrm{V}=\int \xi_{\mathrm{R}}=\begin{pmatrix}
-\frac{1}{2}\dot{A}_{+} & -\frac{1}{2}\dot{A}_{\times} & 0 \\
-\frac{1}{2}\dot{A}_{\times} & +\frac{1}{2}\dot{A}_{+} & 0 \\
0 & 0 & 0
\end{pmatrix}.
\end{equation}
Squaring both sides of the above equation \eqref{V wave} and then taking average gives
\begin{equation}\label{V square average wave}
\overline{\mathrm{V}^2}=\begin{pmatrix}
\frac{1}{4}\left(\dot{A}_{+}^2+\dot{A}_{\times}^2\right) & 0 & 0 \\
0 & \frac{1}{4}\left(\dot{A}_{+}^2+\dot{A}_{\times}^2\right) & 0 \\
0 & 0 & 0
\end{pmatrix}.
\end{equation}

Then inserting \eqref{wave curvature matrix} and \eqref{V square average wave} into \eqref{average evolution equation for X with fluctuations} we obtain the evolution equation for the average path $\overline{\mathbf{X}}$:
\begin{eqnarray}
\frac{d^2\overline{X^1}}{d\tau^2}&=&-\frac{1}{4}\left(\dot{A}_{+}^2+\dot{A}_{\times}^2\right)\overline{X^1},\label{wave attraction 1}\\
\frac{d^2\overline{X^2}}{d\tau^2}&=&-\frac{1}{4}\left(\dot{A}_{+}^2+\dot{A}_{\times}^2\right)\overline{X^2},\label{wave attraction 2}\\
\frac{d^2\overline{X^3}}{d\tau^2}&=&0.\label{wave attraction 3}
\end{eqnarray}

The equations presented above, namely Equations \eqref{wave attraction 1}, \eqref{wave attraction 2} and \eqref{wave attraction 3} for the average evolution $\overline{\mathbf{X}}$, indicate that the plane wave fluctuations generate extra attractions perpendicular to the direction of its propagation. 

It is interesting to look at this example from a different perspective. Let us place the geodesic $\gamma$ in the smooth effective background $g^{(\mathrm{B})}_{ab}$ curved solely by the effective stress energy of the gravitational wave through the effective Einstein equations \eqref{effective Einstein equations for gravity wave} (without the presence of the fluctuations $g^{(\mathrm{F})}_{ab}$). In this scenario, the components of the effective Riemann curvature sourced by the effective stress energy  are given by\footnote{The expressions \eqref{eff R1} and \eqref{eff R2} can be obtained from \eqref{ii} and \eqref{ij} by setting the Weyl curvature components $C=0$.}
\begin{eqnarray}
R^{i(\mathrm{eff})}_{0i0}&=&\frac{4\pi G}{3}\left(t_{00}+2\sum_{k=1}^3t_{kk}-3t_{ii}\right),\label{eff R1}\\
R^{i(\mathrm{eff})}_{0j0}&=&-4\pi Gt_{ij}, \quad i\neq j. \label{eff R2}
\end{eqnarray}

Plugging \eqref{stress energy of gravitational wave} into the above relations \eqref{eff R1} and \eqref{eff R2} gives\footnote{Since for the plane gravitational wave, $A_{+}$ and $A_{\times}$ are small perturbations. Then, to the leading order, the effective stress energy \eqref{stress energy of gravitational wave} in the laboratory frame associated with the TT gauge would be the same as in the geodesic $\gamma$'s local inertial frame $\{(e_0)^a, (e_1)^a, (e_2)^a, (e_3)^a\}$. So we can plug the stress energy \eqref{stress energy of gravitational wave}, which is expressed in the laboratory frame, into the relations \eqref{eff R1} and \eqref{eff R2}, which are expressed in $\gamma$'s local inertial frame, to obtain the expressions \eqref{expression 1} and \eqref{expression 2}.}
\begin{eqnarray}
R^{1(\mathrm{eff})}_{010}&=&R^{2(\mathrm{eff})}_{020}=\frac{1}{4}\left(\dot{A}_{+}^2+\dot{A}_{\times}^2\right),\label{expression 1}\\
R^{3(\mathrm{eff})}_{030}&=&R^{i(\mathrm{eff})}_{0j0}=0, \quad i\neq j. \label{expression 2}
\end{eqnarray}
Expressing the above results in the matrix form, we obtain
\begin{equation}
\mathrm{R}^{(\mathrm{eff})}=\begin{pmatrix}
\frac{1}{4}\left(\dot{A}_{+}^2+\dot{A}_{\times}^2\right) & 0 & 0 \\
0 & \frac{1}{4}\left(\dot{A}_{+}^2+\dot{A}_{\times}^2\right) & 0 \\
0 & 0 & 0
\end{pmatrix}.
\end{equation}

Then from the matrix form geodesic deviation equation \eqref{matrix form} we obtain the evolution $\mathbf{X}^{(\mathrm{eff})}$ in the effective background $g^{(\mathrm{B})}_{ab}$:
\begin{eqnarray}
\frac{d^2 X^1}{d\tau^2}^{(\mathrm{eff})}&=&-\frac{1}{4}\left(\dot{A}_{+}^2+\dot{A}_{\times}^2\right)X^{1(\mathrm{eff})},\label{wave attraction 4}\\
\frac{d^2X^2}{d\tau^2}^{(\mathrm{eff})}&=&-\frac{1}{4}\left(\dot{A}_{+}^2+\dot{A}_{\times}^2\right)X^{1(\mathrm{eff})},\label{wave attraction 5}\\
\frac{d^2X^3}{d\tau^2}^{(\mathrm{eff})}&=&0.\label{wave attraction 6}
\end{eqnarray}

It is important to note that in this example, we have exactly $\overline{\mathrm{V}^2}=\mathrm{R}^{(\mathrm{eff})}$, which implies that the trajectory of the average path $\overline{\mathbf{X}}$ in the spacetime with gravitational wave fluctuations and the trajectory $\mathbf{X}^{(\mathrm{eff})}$ in the effective background $g^{(\mathrm{B})}_{ab}$ sourced by the gravitational wave's effective stress energy follow precisely the same evolution equations \eqref{wave attraction 1}, \eqref{wave attraction 2}, \eqref{wave attraction 3} and \eqref{wave attraction 4}, \eqref{wave attraction 5}, \eqref{wave attraction 6}. This result confirms that the ponderomotive-like attraction is physically equivalent to the attraction produced by the effective stress energy. They both originate from the nonlinearity of gravity.

Given that the ponderomotive-like force is consistently attractive and the derivation holds for generic non-perturbative rapid fluctuations, it follows that, in general, the rapid fluctuations $g_{ab}^{\mathrm{(F)}}$ will always cause the slowly varying background $g_{ab}^{\mathrm{(B)}}$ to curve in a manner that generates attractions. In other words, the effect of the back reaction $t_{ab}$ on the right hand side of the effective Einstein equations \eqref{effective Einstein equations}, arising from the rapid fluctuations $g_{ab}^{\mathrm{(F)}}$, is always attractive.

In fact, the fluctuation induced attraction is a ubiquitous feature of gravity. In order to gain a deeper understanding of the underlying physical mechanism, in the next subsection we demonstrate the existence of the attraction from an alternative perspective.

\subsection{The Extra Shear Attraction}\label{shear attraction}
In general, for any spacetime one can construct the Gaussian normal coordinates (also known as synchronous coordinates) (see, e.g., Wald  1984, \cite{Wald:1984rg} p.42)
\begin{equation}\label{general metric}
ds^2=-dt^2+h_{ij}(t,\mathbf{x})dx^idx^j, \quad i,j=1,2,3.
\end{equation}

In this coordinate system, the world lines $\mathbf{x}=\mathrm{constants}$ are geodesics orthogonal to the hypersurfaces $\Sigma_t$ defined by the equation $t=\mathrm{constant}$. Let's consider a comoving region (the region with fixed spatial coordinates) $\mathcal{U}$ in $\Sigma_t$. The physical volume $V_{\mathcal{U}}$ of the region $\mathcal{U}$ is given by
\begin{equation}\label{volume definition}
V_{\mathcal{U}}(t)=\int_{\mathcal{U}}d^3x\,\sqrt{h} , \quad\text{with}\,\, h=\mathrm{det}(h_{ij}).
\end{equation}
It represents the volume occupied by the free falling test particles enclosed in the region $\mathcal{U}$ whose world lines are orthogonal to $\Sigma_t$ (i.e., the world lines $\mathbf{x}=\mathrm{constants}$, with $\mathbf{x}\in \mathcal{U}$). In the following, we will demonstrate that the Riemann curvature fluctuations $\xi_{R_{abcd}}$ in \eqref{Riemann curvature fluctuation} give rise to extra attraction that consistently tends to reduce the geodesic volume $V_{\mathcal{U}}$.

The Einstein field equations in this coordinate system are equivalent to the following evolution equations for each component $h_{ij}$ of the spatial metric\footnote{Equations \eqref{hij evolution}, \eqref{Hamiltonian constraint} and \eqref{momentum constraint} can be obtained from the equations (20), (21) and (22) of our previous paper \cite{PhysRevD.102.023537}.},
\begin{eqnarray}
\label{hij evolution}
\ddot{h}_{ij}=&-&2R^{(3)}_{ij}-\frac{1}{2}h^{kl}\dot{h}_{ij}\dot{h}_{kl}+h^{kl}\dot{h}_{ik}\dot{h}_{lj}\\
&+&8\pi G\left(T_{00}-h^{kl}T_{kl}\right)h_{ij}+16\pi GT_{ij},\nonumber
\end{eqnarray}
plus the Hamiltonian and momentum constraint equations
\begin{eqnarray}
R^{(3)}+\frac{1}{4}\left(h^{ij}h^{kl}-h^{ik}h^{jl}\right)\dot{h}_{ij}\dot{h}_{kl}&=&16\pi GT_{00},\label{Hamiltonian constraint}\\
D^j\dot{h}_{ij}-D_i\left(h^{jk}\dot{h}_{jk}\right)&=&16\pi GT_{0i},\label{momentum constraint}
\end{eqnarray}
where $R_{ij}^{(3)}$ and $R^{(3)}=h^{ij}R^{(3)}_{ij}$ are the 3-dimensional Ricci tensor and Ricci curvature of the hypersurface $\Sigma_t$, $D_i$ is the derivative operator on the hypersurface $\Sigma_t$ associated with the spatial metric $h_{ij}$.

Define the local scale factor $a(t, \mathbf{x})$ by
\begin{equation}\label{a definition}
h=a^6.
\end{equation}
It describes the relative ``size" of space at each point, which is a generalization of the (global) scale factor $a(t)$ of the usual homogeneous and isotropic FLRW metric \eqref{flrw} in cosmology \cite{PhysRevD.102.023537}.

By combining equations \eqref{hij evolution} and \eqref{Hamiltonian constraint}, we can obtain the evolution equation for $a(t, \mathbf{x})$ \cite{PhysRevD.102.023537}:
\begin{equation}\label{evo}
\ddot{a}=-\left(\frac{4\pi G}{3}\left(T_{00}+h^{ij}T_{ij}\right)+\frac{2\sigma^2}{3}\right)a.
\end{equation}
Here the shear $\sigma^2$ is defined by
\begin{equation}
\sigma^2=\frac{1}{2}\sigma_{ij}\sigma^{ij},
\end{equation}
where $\sigma_{ij}$ is given by
\begin{equation}\label{sigmaij definition}
\sigma_{ij}=\frac{1}{2}\dot{h}_{ij}-\frac{1}{6}h_{ij}h^{kl}\dot{h}_{kl}.
\end{equation}

The shear tensor $\sigma_{ij}$ defined by \eqref{sigmaij definition} is traceless since $h^{ij}\sigma_{ij}=0$. Its evolution is given by\footnote{This equation can be derived from equations (9.2.12), (9.2.13) and (9.2.15) in Wald 1984, \cite{Wald:1984rg} p.218.}
\begin{eqnarray}\label{shear evolution}
\dot{\sigma}_{ij}=&&-2\frac{\dot{a}}{a}\sigma_{ij}-\sigma_{ik}\sigma^k_j+\frac{2}{3}h_{ij}\sigma^2\nonumber\\
&&-C_{0i0j}+4\pi G\left(T_{ij}-\frac{1}{3}h_{ij}h^{kl}T_{kl}\right).
\end{eqnarray}

Since $\sigma_{ij}$ is a ``spatial" tensor, by definition we always have $\sigma^2=\frac{1}{2}\sigma_{ij}\sigma^{ij}\geq 0$. It equals zero if and only if $\sigma_{ij}=0$ for all $i$ and $j$. From the second line of \eqref{shear evolution} we observe that fluctuations in the Weyl curvature $C_{0i0j}$ or in the matter fields stress energy $T_{ij}$ (which give rise to fluctuations in the Ricci curvature) inevitably generate nonzero values for $\sigma_{ij}$. Consequently, we always have $\sigma^2>0$. In other words, since the Riemann curvature is composed of the Weyl curvature and the Ricci curvature (as shown in equation \eqref{Rabcd}), the Riemann curvature fluctuations $\xi_{R_{abcd}}$ in \eqref{Riemann curvature fluctuation} would consistently induce a positive $\sigma^2$.

From \eqref{evo}, it is evident that a positive shear $\sigma^2$ always gives a negative contribution to the relative acceleration $\ddot{a}/a$ of the local scale factor $a(t, \mathbf{x})$. Thus, if $\dot{a}/a>0$, i.e., if locally the space is expanding, the shear tends to slow down the expansion; if $\dot{a}/a<0$, i.e., if locally the space is contracting, the shear tends to speed up the contraction. In other words, the fluctuation induced shear always generates attraction that tends to reduce the geodesic volume $V_{\mathcal{U}}$.

To gain a deeper comprehension of the underlying physical mechanisms responsible for the attractive nature of shear, we can utilize the findings derived in this subsection and apply them to the fluctuations of the plane gravitational wave described by \eqref{plane wave metric 1} and \eqref{plane wave meric 2}, which is propagating in the $z$-direction ($x^3$-direction).

In this specific example, the leading-order metric components are given by:
\begin{eqnarray}
h_{11}&=&1+A_+(t-z), \\
h_{22}&=&1-A_+(t-z), \\
h_{12}&=&h_{21}=A_{\times}(t-z),
\end{eqnarray}
with all other components being zero. By performing direct calculations using the definition \eqref{sigmaij definition} for the shear, we obtain
\begin{equation}
\sigma^2=\frac{1}{4}\left(\dot{A}^2_{+}+\dot{A}^2_{\times}\right).
\end{equation}

In the case of a gravitational wave, there is no matter fields stress energy $T_{ij}$ and then the evolution equation \eqref{evo} for the local scale factor $a(t, \mathbf{x})$ becomes
\begin{equation}\label{wave attraction for a}
\ddot{a}=-\frac{2}{3}\times\frac{1}{4}\left(\dot{A}^2_{+}+\dot{A}^2_{\times}\right)a.
\end{equation}
This equation describes the dynamics of $a(t, \mathbf{x})$ in response to the gravitational wave fluctuations, where the attraction is proportional to the square of the time derivatives of the wave amplitudes $A_+$ and $A_{\times}$.

It is intriguing to observe that the above equation \eqref{wave attraction for a} corresponds exactly to the ``average" of the three equations \eqref{wave attraction 1}, \eqref{wave attraction 2}, \eqref{wave attraction 3} for $\overline{\mathbf{X}}$ or the three equations \eqref{wave attraction 4}, \eqref{wave attraction 5}, \eqref{wave attraction 6} for $\mathbf{X}^{(\mathrm{eff})}$. The presence of the factor $\frac{2}{3}$ can be easily understood since, for this plane gravitational wave, the attraction only occur in the two perpendicular directions, namely, $x$ and $y$, with respect to the direction of propagation, $z$. This result demonstrates that the extra shear attraction share the same physical origin as the extra ponderomotive-like tidal attraction and the attraction generated by the effective stress energy. They all arise from the nonlinear nature of gravity. Physically, the extra ponderomotive-like tidal attraction between neighboring geodesics results in a tendency for the geodesic volume $V_{\mathcal{U}}$ to shrink. The shear attraction represents the average of the ponderomotive-like tidal attraction between geodesics over the three spatial directions.

\section{The Quantum Fluctuations in Riemann Curvature}\label{quantum fluctuations}
The uncertainty principle implies that the quantum fluctuations in the spacetime curvature are inescapable. They are superposed on and coexist with the large-scale, slowly varying curvature predicted by classical deterministic general relativity (Wheeler 1973, \cite{Misner:1973prb} p.1192-1193). In this section, we investigate the properties of these fluctuations.

\subsection{Two Sources of the Curvature Fluctuations}
The Riemann tensor can be expressed in terms of the Weyl tensor and the Ricci tensor as follows:
\begin{equation}\label{Rabcd}
R_{abcd}=C_{abcd}+g_{a[c}R_{d]b}-g_{b[c}R_{d]a}-\frac{1}{3}Rg_{a[c}g_{d]b}.
\end{equation}

The Ricci tensor is the trace part of the Riemann tensor. It is determined by the matter distribution point-to-point through the Einstein field equations:
\begin{equation}\label{EFE}
R_{ab}=8\pi G\left(T_{ab}-\frac{1}{2}Tg_{ab}\right),
\end{equation}
where $T=g^{ab}T_{ab}$ is the trace of the matter fields stress-energy tensor.

The Weyl tensor is the trace-free part of the Riemann tensor. It is not determined locally by matter distribution. However, it cannot be entirely arbitrary as the Riemann tensor must satisfy the Bianchi identity:
\begin{equation}\label{Bianchi}
\nabla_{[a}R_{bc]de}=0.
\end{equation}
In terms of the Weyl and the Ricci tensors, \eqref{Bianchi} becomes
\begin{equation}\label{weyl bian}
\nabla^aC_{abcd}=\nabla_{[c}R_{d]b}+\frac{1}{6}g_{b[c}\nabla_{d]}R.
\end{equation}
Substituting the Einstein equations \eqref{EFE} into \eqref{weyl bian} we have
\begin{equation}\label{eom of weyl}
\nabla^aC_{abcd}=8\pi G\left(\nabla_{[c}T_{d]b}+\frac{1}{3}g_{b[c}\nabla_{d]}T\right).
\end{equation}

Thus, the Weyl tensor and the matter fields stress energy tensor are related by the above first-order differential equations \eqref{eom of weyl}. Given some specified stress-energy distribution, there is still freedom in the choice of the Weyl curvature (see, e.g., Carroll 2019, \cite{Carroll:2004st} p.169). For convenience, we express the solution to \eqref{eom of weyl} in the form
\begin{equation}\label{weyl decomposition}
C_{abcd}=C^{(\mathrm{free})}_{abcd}+C^{(\mathrm{mat})}_{abcd},
\end{equation}
where
\begin{eqnarray}
\nabla^aC^{\mathrm{(free)}}_{abcd}&=&0,\label{free weyl}\\
\nabla^aC^{\mathrm{(mat)}}_{abcd}&=&8\pi G\left(\nabla_{[c}T_{d]b}+\frac{1}{3}g_{b[c}\nabla_{d]}T\right).\label{matter weyl}
\end{eqnarray}
In some sense $C^{\mathrm{(free)}}_{abcd}$ represents the free part of Weyl curvature which is not sourced by the matter distribution\footnote{Strictly speaking, $C^{\mathrm{(free)}}_{abcd}$ is not completely free, since the derivative operator $\nabla^a$ in \eqref{free weyl} is affected by the curvature produced by the matter distribution.}, while $C^{(\mathrm{mat})}_{abcd}$ represents the part of Weyl curvature at a point that is sourced by the matter distribution at other points (see, e.g., Hawking and Ellis 1973, \cite{Hawking:1973uf} p.85).

Therefore, in a complete theory of quantum gravity, it is expected that the Riemann curvature fluctuations arise from two sources: (i) the passive fluctuations in $R_{ab}$ and $C^{(\mathrm{mat})}_{abcd}$ which are sourced by the fluctuations of the matter fields through \eqref{EFE} and \eqref{eom of weyl}, respectively; (ii) the active fluctuations in $C^{(\mathrm{free})}_{abcd}$ due to the quantum nature of gravity itself \cite{Ford:2000vm}.

\subsection{The Curvature Matrix $\mathrm{R}$}
In the matrix form of the geodesic deviation equation \eqref{matrix form}, each element of the curvature matrix $\mathrm{R}$ is the component $R^i_{0j0}$ of the Riemann curvature tensor $R^a_{bcd}$ in the local inertial frame $\{(e_0)^a, (e_1)^a, (e_2)^a, (e_3)^a\}$ of the geodesic $\gamma$. In this frame, \eqref{Rabcd} becomes
\begin{equation}\label{ri0j00}
R^i_{0j0}=R_{0i0j}=C_{0i0j}-\frac{1}{2}R_{ij}+\frac{1}{2}\delta_{ij}R_{00}+\frac{1}{6}R\delta_{ij}.
\end{equation}

Plugging the Einstein field equations \eqref{EFE} into \eqref{ri0j00} we obtain that the components $R^i_{0j0}$ are related to the matter distribution through the following relations
\begin{eqnarray}
&&R^i_{0i0}=C_{0i0i}+\frac{4\pi G}{3}\left(T_{00}+2\sum_{k=1}^3T_{kk}-3T_{ii}\right) \label{ii}\\
&&R^i_{0j0}=C_{0i0j}-4\pi GT_{ij}, \quad i\neq j.  \label{ij}
\end{eqnarray}

The above equations \eqref{ii} and \eqref{ij} can be written in the matrix form
\begin{equation}\label{curvature matrix}
\mathrm{R}=\mathrm{C}+\mathrm{M},
\end{equation}
where the matrix $\mathrm{C}$, which represents the Weyl curvature contribution, is defined by
\begin{equation}\label{Weyl curvature matrix}
\mathrm{C}=\begin{pmatrix}
C_{0101} & C_{0102} & C_{0103} \\
C_{0201} & C_{0202} & C_{0203} \\
C_{0301} & C_{0302} & C_{0303}
\end{pmatrix},
\end{equation}
and the matrix $\mathrm{M}$, which represents the matter fields contribution, is defined by
\begin{widetext}
\begin{equation}\label{M matrix}
\mathrm{M}=\frac{4\pi G}{3}\begin{pmatrix}
T_{00}-T_{11}+2T_{22}+2T_{33} & -3T_{12} & -3T_{13} \\
 -3T_{21}& T_{00}-T_{22}+2T_{11}+2T_{33} & -3T_{23} \\
-3T_{31} &-3T_{32} & T_{00}-T_{33}+2T_{11}+2T_{22}
\end{pmatrix}.
\end{equation}
\end{widetext}

By Eq.\eqref{weyl decomposition}, the matrix $\mathrm{C}$ can be further decomposed as
\begin{equation}\label{weyl matrix decomposition}
\mathrm{C}=\mathrm{C}^{(\mathrm{free})}+\mathrm{C}^{(\mathrm{mat})},
\end{equation}
where the element of the matrix $\mathrm{C}^{(\mathrm{free})}$ is $\mathrm{C}^{(\mathrm{free})}_{ij}=C^{(\mathrm{free})}_{0i0j}$, the element of the matrix $\mathrm{C}^{(\mathrm{mat})}$ is $\mathrm{C}^{(\mathrm{mat})}_{ij}=C^{(\mathrm{mat})}_{0i0j}$.

Then the Riemann curvature matrix $\mathrm{R}$ along the geodesic $\gamma$ parameterized by $\tau$ can be written as
\begin{equation}
\mathrm{R}(\tau)=\mathrm{M}(\tau)+\mathrm{C}^{(\mathrm{mat})}(\tau)+\mathrm{C}^{(\mathrm{free})}(\tau).
\end{equation}

Correspondingly, the fluctuation $\xi_{\mathrm{R}}$ in $\mathrm{R}$ can be decomposed as the sum
\begin{equation}\label{R fluctuation}
\xi_{\mathrm{R}}(\tau)=\xi_{\mathrm{M}}(\tau)+\xi_{\mathrm{C}^{(\mathrm{mat})}}(\tau)+\xi_{\mathrm{C}^{(\mathrm{free})}}(\tau),
\end{equation}
where the first two terms $\xi_{\mathrm{M}}$ and $\xi_{\mathrm{C}^{(\mathrm{mat})}}$ represent the passive fluctuations sourced by the matter fields, the last term $\xi_{\mathrm{C}^{(\mathrm{free})}}$ represents the active fluctuations due to the quantum nature of gravity itself. In the next two subsections, we investigate the properties of these fluctuations.

\subsection{Passive Fluctuations from the Matter Fields}\label{passive fluctuation}
Quantum fluctuation is a ubiquitous feature of matter fields which has been experimentally verified by various phenomena like the Lamb shift \cite{PhysRev.74.1157}, the Casimir effect \cite{Klimchitskaya:2006rw}, the spontaneous emission \cite{1984AmJPh..52..340M}, the anomalous magnetic moment of the electron \cite{PhysRev.72.1256.2}, among others. In particular, there exist large zero-point fluctuations in the stress-energy tensor of the quantum matter fields.

As an illustrative example\footnote{On dimensional grounds, the results we obtained from this example, like the leading order dependence on the matter fields high energy cutoff $\Lambda_{\mathrm{mat}}$, also apply to other fields in curved spacetime.}, let us consider a quantized free real scalar field in Minkowski spacetime
\begin{equation}
\phi(t, \mathbf{x})=\int\frac{d^3k}{(2\pi)^{3/2}}\frac{1}{\sqrt{2\omega}}\left(a_{\mathbf{k}}e^{-i(\omega t-\mathbf{k}\cdot\mathbf{x})}+a_{\mathbf{k}}^{\dag}e^{+i(\omega t-\mathbf{k}\cdot\mathbf{x})}\right),
\end{equation}
where $\omega=\sqrt{\mathbf{k}^2+m^2}$.

The Hamiltonian is the integration of the energy density over the whole space
\begin{equation}\label{Hamiltonian}
\mathrm{H}=\int d^3x\, T_{00}=\frac{1}{2}\int d^3k\omega\left(a_{\mathbf{k}}a_{\mathbf{k}}^{\dag}+a_{\mathbf{k}}^{\dag}a_{\mathbf{k}}\right),
\end{equation}
where
\begin{widetext}
\begin{eqnarray}
T_{00}\left(t, \mathbf{x}\right)&=&\frac{1}{2}\left(\left(\frac{\partial\phi}{\partial t}\right)^2+\left(\nabla\phi\right)^2+m^2\phi^2\right)\label{T00 definition}\\
&=&\frac{1}{2}\int\frac{d^3kd^3k'}{\left(2\pi\right)^3}\Bigg(\frac{\left(\omega\omega'+\mathbf{k}\cdot\mathbf{k}'+m^2\right)}{2\sqrt{\omega\omega'}}\left(a_{\mathbf{k}}a_{\mathbf{k}'}^{\dag}e^{-i\left[\left(\omega-\omega'\right)t-\left(\mathbf{k}-\mathbf{k}'\right)\cdot\mathbf{x}\right]}+a_{\mathbf{k}}^{\dag}a_{\mathbf{k}'}e^{+i\left[\left(\omega-\omega'\right)t-\left(\mathbf{k}-\mathbf{k}'\right)\cdot\mathbf{x}\right]}\right)\nonumber\\
&&\quad\quad\quad\quad\quad -\frac{\left(\omega\omega'+\mathbf{k}\cdot\mathbf{k}'-m^2\right)}{2\sqrt{\omega\omega'}}\left(a_{\mathbf{k}}a_{\mathbf{k}'}e^{-i\left[\left(\omega+\omega'\right)t-\left(\mathbf{k}+\mathbf{k}'\right)\cdot\mathbf{x}\right]}
+a_{\mathbf{k}}^{\dag}a_{\mathbf{k}'}^{\dag}e^{+i\left[\left(\omega+\omega'\right)t-\left(\mathbf{k}+\mathbf{k}'\right)\cdot\mathbf{x}\right]}\right)\Bigg).
\end{eqnarray}
\end{widetext}

It is important to note that although the Hamiltonian operator $H$ is independent of time, both the field operator $\phi$ and the energy density operator $T_{00}$ are time-dependent. In fact, since $\phi$ and $T_{00}$ do not commute with $H$, we have the Heisenberg equations of motion
\begin{eqnarray}
\frac{\partial\phi}{\partial t}&=&-i[\phi, H]\neq 0,\label{Heisenberge for phi}\\
\frac{\partial T_{00}}{\partial t}&=&-i[T_{00}, H]\neq 0,\label{Heisenberge for T00}
\end{eqnarray}
which imply that both the field $\phi$ and the energy density $T_{00}$ vary with time.

For any eigenstate $|\Psi\rangle$ of the Hamiltonian (stationary state) we have the expectation values of the field $\phi$ and the energy density $T_{00}$
\begin{eqnarray}
\langle\Psi|\phi|\Psi\rangle&\equiv&\mathrm{constant},\\
\langle\Psi|T_{00}|\Psi\rangle&\equiv&\mathrm{constant},
\end{eqnarray}
because the expectation values of their rates of change
\begin{eqnarray}
\left\langle\Psi\left|\frac{\partial\phi}{\partial t}\right|\Psi\right\rangle&=&0,\label{a}\\
\left\langle\Psi\left|\frac{\partial T_{00}}{\partial t}\right|\Psi\right\rangle&=&0.\label{b}
\end{eqnarray}
However, the expectation values of the square of their rates of change
\begin{eqnarray}
\left\langle\Psi\left|\left(\frac{\partial\phi}{\partial t}\right)^2\right|\Psi\right\rangle&\neq&0,\label{c}\\
\left\langle\Psi\left|\left(\frac{\partial T_{00}}{\partial t}\right)^2\right|\Psi\right\rangle&\neq&0.\label{d}
\end{eqnarray}
This implies that $\frac{\partial\phi}{\partial t}$ and $\frac{\partial T_{00}}{\partial t}$ take random values about zero, indicating that both $\phi$ and $T_{00}$ constantly fluctuate.

In particular, consider the vacuum state $|0\rangle$ which is defined as $a_{\mathbf{k}}|0\rangle=0$, for all $\mathbf{k}$. It is an eigenstate of the total Hamiltonian operator $H$, but not an an eigenstate of the field operator $\phi$ and the energy density operator $T_{00}$. From the above analysis, both $\phi$ and $T_{00}$ exhibit zero-point fluctuations in the vacuum state.

Furthermore, from the definition \eqref{T00 definition} for $T_{00}$, the zero-point fluctuations in $\phi$ lead to nonzero expectation value of the zero-point energy density $\left\langle 0| T_{00}|0\right\rangle$. It is well known that this value is divergent if one tries to evaluate the field at a point or equivalently to sum the field modes to arbitrarily high energies. However, field values cannot be observed at a point; in realistic experiments, only averages of field values over a finite region of space and time can be measured. 

If we average the energy density $T_{00}$ over a spacetime region of size $L$, or equivalently, we impose a high energy cutoff $\Lambda_{\mathrm{mat}}=1/L$, then to leading order, the expectation value of the zero-point energy density goes as
\begin{equation}\label{magnitude of the vacuum energy density}
\left\langle 0| T_{00}|0\right\rangle\sim\frac{1}{L^4}=\Lambda_{\mathrm{mat}}^4.
\end{equation}
This result could also have been guessed from dimensional analysis.

Meanwhile, the density of the zero-point energy, originating from the zero-point fluctuations of the field $\phi$, also undergoes fluctuations. The magnitude of the fluctuations about the average value $\left\langle 0| T_{00}|0\right\rangle$ is comparable to the average itself. By direct calculations (or by simple dimensional analysis), we obtain that, to leading order, the magnitude of the fluctuations goes as
\begin{equation}
\Delta T_{00}\sim\sqrt{\left\langle\left(T_{00}-\left\langle T_{00}\right\rangle\right)^2\right\rangle}\sim\left\langle T_{00}\right\rangle\sim\Lambda_{\mathrm{mat}}^4.
\end{equation}

Furthermore, the magnitude of the rate of change of $T_{00}$ goes as
\begin{equation}
\sqrt{\left\langle\left(\partial T_{00}/\partial t\right)^2\right\rangle} \sim\Lambda_{\mathrm{mat}}^5.
\end{equation}
Consequently, $T_{00}$ fluctuates on the time scale
\begin{equation}
\Delta t\sim \frac{\Delta T_{00}}{\sqrt{\left\langle\left(\partial T_{00}/\partial t\right)^2\right\rangle}}\sim 1/\Lambda_{\mathrm{mat}}.
\end{equation}
This result could also have been guessed from energy time uncertainty principle.

The above analysis for $T_{00}$ can also be applied to the other components $T_{\mu\nu}$ of the stress energy tensor $T_{ab}$. In general, we have, at leading order, the magnitude of the fluctuations in $T_{\mu\nu}$ goes as
\begin{equation}\label{delta tmunu}
\Delta T_{\mu\nu}\sim\sqrt{\left\langle\left(T_{\mu\nu}-\left\langle T_{\mu\nu}\right\rangle\right)^2\right\rangle}\sim\Lambda_{\mathrm{mat}}^4,
\end{equation}
the magnitude of the rate of change of $T_{\mu\nu}$ goes as
\begin{equation}
\sqrt{\left\langle\left(\partial T_{\mu\nu}/\partial t\right)^2\right\rangle}\sim\Lambda_{\mathrm{mat}}^5,
\end{equation}
and $T_{\mu\nu}$ fluctuates on the time scale
\begin{equation}\label{time scale tmunu}
\Delta t\sim \frac{\Delta T_{\mu\nu}}{\sqrt{\left\langle\left(\partial T_{\mu\nu}/\partial t\right)^2\right\rangle}}\sim 1/\Lambda_{\mathrm{mat}}.
\end{equation}

In the local inertial frame $\{(e_0)^a, (e_1)^a, (e_2)^a, (e_3)^a\}$ of $\gamma$, the spacetime is approximately flat within a small neighborhood of $\gamma$. Therefore, the above flat spacetime results also apply to the matrix $\mathrm{M}(\tau)$, which is defined along the trajectory of $\gamma$. Since each element $\mathrm{M}_{ij}(\tau)$ of the matrix $\mathrm{M}(\tau)$ is a linear combination of the components of $T_{\mu\nu}$, we have the magnitude of the fluctuations in $\mathrm{M}_{ij}$ goes as
\begin{equation}\label{M fluctuation}
\xi_{\mathrm{M}_{ij}}\sim G\Delta T_{\mu\nu} \sim G\Lambda_{\mathrm{mat}}^4,
\end{equation}
the magnitude of the rate of change of $\mathrm{M}_{ij}$ goes as
\begin{equation}
\sqrt{\left\langle\left(\partial \mathrm{M}_{ij}/\partial\tau\right)^2\right\rangle}\sim G\Lambda_{\mathrm{mat}}^5,
\end{equation}
and $\mathrm{M}_{ij}(\tau)$ fluctuates along $\gamma$ on the time scale
\begin{equation}\label{time scale 1}
T\sim \frac{\xi_{\mathrm{M}_{ij}}}{\sqrt{\left\langle\left(\partial \mathrm{M}_{ij}/\partial\tau\right)^2\right\rangle}}\sim 1/\Lambda_{\mathrm{mat}}.
\end{equation}

Likewise, we have the magnitude of the fluctuations in $\mathrm{C}^{(\mathrm{mat})}_{ij}$ goes as
\begin{equation}\label{C mat fluctuation}
\xi_{\mathrm{C}^{(\mathrm{mat})}_{ij}}\sim G\Lambda_{\mathrm{mat}}^4,
\end{equation}
and $\mathrm{C}^{(\mathrm{mat})}_{ij}(\tau)$ fluctuates along $\gamma$ on the time scale
\begin{equation}\label{time scale 2}
T\sim 1/\Lambda_{\mathrm{mat}}.
\end{equation}

\subsection{Active Fluctuations of the Weyl Curvature}\label{active fluctuations}
Wheeler has emphasized the significance of quantum fluctuations in spacetime curvature at distances comparable to Planck's length $L_p=\sqrt{G}$. His proposition portrays the gravitational vacuum as a foamlike structure vigorously resonating on small scales among various configurations with different geometries and topologies. Base on dimensional analysis and analogies with quantum electrodynamics, he arrived at the conclusions that, in a region of dimension $L$, there will occur fluctuations in the typical components of the metric $g$ of the order (see e.g. Wheeler 1973, \cite{Misner:1973prb} p.1192-1193)
\begin{equation}\label{wheeler 1}
\Delta g\sim\frac{L_p}{L},
\end{equation}
fluctuations in the affine connection (first order derivatives of the metric) $\Gamma$ of the order
\begin{equation}\label{wheeler 2}
\Delta \Gamma\sim\frac{\Delta g}{L}\sim\frac{L_p}{L^2},
\end{equation} 
and fluctuations in the curvature $R$ of the order
\begin{equation}\label{wheeler 3}
\Delta R\sim\frac{\Delta g}{L^2}\sim\frac{L_p}{L^3}.
\end{equation}

Equivalently, if we impose a high energy cutoff $\Lambda_{\mathrm{grav}}=1/L$ on the gravitational field, then the fluctuations in the above geometric quantities go as
\begin{eqnarray}
\Delta g&\sim& \sqrt{G}\Lambda_{\mathrm{grav}},\\
\Delta \Gamma&\sim&\sqrt{G}\Lambda_{\mathrm{grav}}^2,\label{metric derivative fluctuation}\\
\Delta R&\sim& \sqrt{G}\Lambda_{\mathrm{grav}}^3.\label{wheeler cuvature estimation}
\end{eqnarray}

These fluctuations of the gravitational vacuum correspond to the active fluctuations in the free part of the Weyl curvature $C^{\mathrm{(free)}}_{abcd}$. Thus, we have the magnitude of the fluctuations in each element $\mathrm{C}^{(\mathrm{free})}_{ij}$ of the matrix $\mathrm{C}^{(\mathrm{free})}$ goes as
\begin{equation}\label{C free fluctuation}
\xi_{\mathrm{C}_{ij}^{(\mathrm{free})}}\sim\sqrt{G}\Lambda_{\mathrm{grav}}^3.
\end{equation}

As the affine connection is the first derivative of the metric, the magnitude of the rate of change of the metric should be on the same order of the fluctuations in the affine connection
\begin{equation}
\sqrt{\left\langle\left(\partial g/\partial t\right)^2\right\rangle}\sim\Delta\Gamma\sim\sqrt{G}\Lambda_{\mathrm{grav}}^2.
\end{equation}
Consequently, the time scale of variation of the metric $g$ is on the order
\begin{equation}
\Delta t\sim\frac{\Delta g}{\sqrt{\left\langle\left(\partial g/\partial t\right)^2\right\rangle}}=1/\Lambda_{\mathrm{grav}}.
\end{equation}

Since the time scale of variation for each element $\mathrm{C}^{(\mathrm{free})}_{ij}$ of the matrix $\mathrm{C}^{(\mathrm{free})}$ is the same as the time scale of variation of the metric $g$, it follows that these elements also fluctuate along $\gamma$ on the time scale
\begin{equation}\label{time scale 3}
T\sim 1/\Lambda_{\mathrm{grav}}.
\end{equation}

\subsection{The Cutoffs $\Lambda_{\mathrm{mat}}$ and $\Lambda_{\mathrm{grav}}$}\label{cutoff}
Quantum field theory is an effective description of matter at low energies. It might be a low energy approximation to a more fundamental theory that may not be a field theory at all \cite{Weinberg:2021exr}. The cutoff $\Lambda_{\mathrm{mat}}$ represents the energy scale past which we no longer trust the field theory description of matter. We do not know the exact scale $\Lambda_{\mathrm{mat}}$ at which the field description will fail. However, since quantum field theory description of matter is defined on classical fixed spacetime, it is generally expected that $\Lambda_{\mathrm{mat}}$ will be lower than the Planck energy $E_p=\frac{1}{\sqrt{G}}$, the scale at which quantum gravity effect becomes crucial and the classical description of spacetime breaks down. For example, it has been proposed in \cite{PhysRevD.65.105013} that gravitational stability of the vacuum sets a limit $E_p/\mathcal{N}^{1/4}$ on the shortest scale of any quantum field theory compatible with gravity, where $\mathcal{N}$ is the number of matter fields. Moreover, it is more likely that the field theory fails much more earlier. For example, it has been proposed in \cite{Carmona:2000gd} that the upper limit of the domain of validity of the quantum field theory description of matter is around $100$ $\mathrm{TeV}$. Therefore, we expect that the cutoff of the matter fields
\begin{equation}\label{matter cutoff}
\Lambda_{\mathrm{mat}}\leq E_p,
\end{equation}
or, more likely,
\begin{equation}\label{matter cutoff 1}
\Lambda_{\mathrm{mat}}\ll E_p.
\end{equation}

General relativity is a classical description of spacetime and gravity. The quantum effect of gravity starts to become crucial at the Planck scale and the classical description is expected to break down at that scale. According to Wheeler's conclusions \eqref{wheeler 1}, \eqref{wheeler 2} and \eqref{wheeler 3}, quantum fluctuations in the geometry are superposed on and coexist with the classical deterministic geometry described by general relativity. The classical dynamics of spacetime geometry is expected to be replaced by quantum geometrodynamics at and above the Planck energy scale.

We do not know whether Wheeler's quantum geometrodynamical description of spacetime is valid for all scales above the Planck energy scale or there exists a finite cutoff $\Lambda_{\mathrm{grav}}$ past which the quantum geometrical description is no longer valid and a new theory of gravity which does not rely on the notion of spacetime geometry is needed. However, in analogue with other three fundamental interactions, the Compton wavelength of a particle is the length scale at which quantum field theory starts to become crucial for its accurate description and we know that this description is valid to distance scales many orders of magnitude smaller than the Compton wavelength. Thus we may expect that the quantum geometrodynamcial description of spacetime is also valid to distances scales many orders of magnitude smaller than the Planck length. Therefore, we expect that the cutoff of gravity
\begin{equation}
\Lambda_{\mathrm{grav}}\geq E_p,
\end{equation}
or, more likely
\begin{equation}\label{gravity cutoff}
\Lambda_{\mathrm{grav}}\gg E_p.
\end{equation}

\section{The Extra Attraction Generated by the Quantum Fluctuations}\label{extra attraction generated by quantum fluctuations}
Next we investigate the properties of the extra attraction induced by quantum fluctuations.

\subsection{The Quantum Tidal Attraction}
In this subsection, we investigate the extra tidal attraction generated by quantum fluctuations in the Riemann curvature by applying the properties of the quantum fluctuations we obtained in Sec.\ref{quantum fluctuations} to the results we obtained in Sec.\ref{attraction}.

From the Einstein field equations \eqref{EFE}, we have, in the vacuum of matter fields, the typical value of the average background Ricci curvature $\overline{R}_{\mu\nu}$ is of the order
\begin{equation}
\overline{R}_{\mu\nu}\sim G\left\langle T_{00}\right\rangle\sim G\Lambda_{\mathrm{mat}}^4.
\end{equation}
Meanwhile, we have, in the vacuum of gravitational field, the average background Weyl curvature $\overline{\mathrm{C}}^{\mu}_{\nu\lambda\rho}$ is of the order
\begin{equation}
\overline{\mathrm{C}}^{\mu}_{\nu\lambda\rho}\sim 0.
\end{equation}
Thus, according to \eqref{Rabcd}, the typical value of the average background Riemann curvature $\overline{R}^{\mu}_{\nu\lambda\rho}$ is of the same order as the Ricci curvature:
\begin{equation}
\overline{R}^{\mu}_{\nu\lambda\rho}\sim\overline{R}_{\mu\nu}\sim G\left\langle T_{00}\right\rangle\sim G\Lambda_{\mathrm{mat}}^4.
\end{equation}
Then the time scale of variation of the background spacetime metric is of the order
\begin{equation}\label{time scale 5}
1/|\overline{R}^{\mu}_{\nu\lambda\rho}|^{1/2}\sim\frac{1}{\sqrt{G}\Lambda_{\mathrm{mat}}^2}.
\end{equation}

In contrast, according to \eqref{time scale 1}, \eqref{time scale 2} and \eqref{time scale 3}, we have, each element $\mathrm{R}_{ij}$ of the curvature matrix $\mathrm{R}$ fluctuates along $\gamma$ on the time scales
\begin{equation}\label{time scale 4}
T\sim 1/\Lambda_{\mathrm{mat}} \,\,\text{and}\,\, 1/\Lambda_{\mathrm{grav}}.
\end{equation}

In the regime given by \eqref{matter cutoff 1} and \eqref{gravity cutoff}, we have the relation $\left(\sqrt{G}\Lambda_{\mathrm{mat}}\right)^2\ll\sqrt{G}\Lambda_{\mathrm{mat}}\ll1$ so that $\sqrt{G}\Lambda_{\mathrm{mat}}^2\ll\Lambda_{\mathrm{mat}}\ll\Lambda_{\mathrm{grav}}$ and thus
\begin{equation}\label{time scale comparing}
1/\Lambda_{\mathrm{grav}}\ll 1/\Lambda_{\mathrm{mat}}\ll\frac{1}{\sqrt{G}\Lambda_{\mathrm{mat}}^2}.
\end{equation}
Then from \eqref{time scale 5}, \eqref{time scale 4} and \eqref{time scale comparing}, we have the time scale of the fluctuations in the curvature is much smaller than time scale of variation of the background spacetime metric. Hence, the assumption \eqref{time scale of variation} of relatively rapid curvature fluctuation is satisfied, and the results obtained in Sec.\ref{attraction} are applicable to the quantum fluctuations discussed in Sec.\ref{quantum fluctuations}.

Substituting \eqref{M fluctuation}, \eqref{C mat fluctuation} and \eqref{C free fluctuation} into \eqref{R fluctuation} we obtain that the magnitude of the fluctuations in each element $\mathrm{R}_{ij}$ of the curvature matrix $\mathrm{R}$ goes as
\begin{equation}\label{Rij fluctuation}
\xi_{\mathrm{R}_{ij}}\sim\alpha_{ij}\sqrt{G}\Lambda_{\mathrm{grav}}^3+\beta_{ij} G\Lambda_{\mathrm{mat}}^4,
\end{equation}
where the values of the dimensionless parameters $\alpha_{ij}$ and $\beta_{ij}$ (and the parameters $\alpha'_{ij}$, $\beta'_{ij}$ in \eqref{Vij fluctuation}, $\alpha'$, $\beta'$ in \eqref{vsquare magnitude}) depend on the specific details of the quantum fluctuations in the matter fields and in the free part of the Weyl curvature.

Using \eqref{Rij fluctuation} and \eqref{time scale 4} we have the magnitude of each element $\mathrm{V}_{ij}$ of the matrix $\mathrm{V}$ defined by \eqref{V definition} goes as
\begin{equation}\label{Vij fluctuation}
\mathrm{V}_{ij}=\int^{\tau}\xi_{\mathrm{R}_{ij}}d\tau'\sim \alpha_{ij}'\sqrt{G}\Lambda_{\mathrm{grav}}^2+\beta_{ij}'G\Lambda_{\mathrm{mat}}^3.
\end{equation}

In the free falling observer $\gamma$'s local inertial frame $\{(e_0)^a, (e_1)^a, (e_2)^a, (e_3)^a\}$, the quantum fluctuations in $\mathrm{R}$ have no preferred directions in the matter fields vacuum and the gravitational vacuum. Hence, the assumed properties \eqref{Vij properties1} and \eqref{Vij properties2} are satisfied, and we can plug \eqref{Vij fluctuation} into \eqref{vsquare definition} to obtain the strength of the extra attraction
\begin{equation}\label{vsquare magnitude}
v^2\sim \alpha' G\Lambda_{\mathrm{grav}}^4+\beta' G^2\Lambda_{\mathrm{mat}}^6.
\end{equation}

The first term $\alpha' G\Lambda_{\mathrm{grav}}^4$ in \eqref{vsquare magnitude} comes from the zero-point fluctuations of the gravitational vacuum, the second term $\beta' G^2\Lambda_{\mathrm{mat}}^4$ comes from the zero-point fluctuations of the matter fields vacuum. From \eqref{matter cutoff 1} and \eqref{gravity cutoff}, we have
\begin{equation}\label{inequality}
G^2\Lambda_{\mathrm{mat}}^6\ll G\Lambda_{\mathrm{mat}}^4\ll G\Lambda_{\mathrm{grav}}^4.
\end{equation}
So the contribution to the extra tidal attraction from the zero-point fluctuations of the gravitational vacuum is dominant and we have, to leading order,
\begin{equation}\label{vsquare estimation}
v^2\propto G\Lambda_{\mathrm{grav}}^4.
\end{equation}
Then from \eqref{Extra force} we have, to leading order, the extra tidal attraction
\begin{equation}
\mathbf{A}_{\mathrm{extra}}\propto -G\Lambda_{\mathrm{grav}}^4\overline{\mathbf{X}}.
\end{equation}

\subsection{The Quantum Shear Attraction}
In this subsection, we investigate the extra shear attraction generated by quantum fluctuations in the Riemann curvature by applying the properties of the quantum fluctuations we obtained in Sec.\ref{quantum fluctuations} to the results we obtained in Sec.\ref{shear attraction}.

From \eqref{wheeler cuvature estimation}, we have the magnitude of the active fluctuations in the free part of the Weyl curvature $C_{0i0j}^{(\mathrm{free})}$ goes as
\begin{equation}\label{fluctuation c free}
\xi_{C_{0i0j}^{(\mathrm{free})}}\sim \sqrt{G}\Lambda_{\mathrm{grav}}^3.
\end{equation}
It fluctuates on the time scale
\begin{equation}\label{time grav}
T\sim 1/\Lambda_{\mathrm{grav}}.
\end{equation}

From \eqref{matter weyl} and \eqref{delta tmunu}, we have the magnitude of the passive fluctuations in the part of Weyl curvature $C_{0i0j}^{(\mathrm{mat})}$ sourced by the matter fields fluctuations goes as
\begin{equation}\label{fluctuation c mat}
\xi_{C_{0i0j}^{(\mathrm{mat})}}\sim G\Lambda_{\mathrm{mat}}^4.
\end{equation}
It fluctuates on the time scale
\begin{equation}\label{time mat}
T\sim 1/\Lambda_{\mathrm{mat}}.
\end{equation}

Integrating \eqref{shear evolution} and using \eqref{weyl decomposition}, \eqref{fluctuation c free}, \eqref{time grav}, \eqref{fluctuation c mat}, \eqref{time mat}, \eqref{delta tmunu} and \eqref{time scale tmunu}, we obtain that the magnitude of the fluctuations in the shear tensor goes as
\begin{equation}\label{magnitude of shear tensor fluctuations}
\sigma_{ij}\sim \tilde{\alpha}_{ij}\sqrt{G}\Lambda_{\mathrm{grav}}^2+\tilde{\beta}_{ij} G\Lambda_{\mathrm{mat}}^3,
\end{equation}
where the values of the dimensionless parameters $\tilde{\alpha}_{ij}$ and $\tilde{\beta}_{ij}$ (and the parameters $\tilde{\alpha}, \tilde{\beta}$ in \eqref{magnitude of shear}) depend on the specific details of the curvature fluctuations.

From the above estimation \eqref{magnitude of shear tensor fluctuations} we obtain that the magnitude of the shear goes as
\begin{equation}\label{magnitude of shear}
\sigma^2\sim\tilde{\alpha}G\Lambda_{\mathrm{grav}}^4+\tilde{\beta}G^2\Lambda_{\mathrm{mat}}^6.
\end{equation}
It takes the same form as the magnitude of $v^2$ given by \eqref{vsquare magnitude}. This result is just what we expected, since our analysis in previous sections has shown that the ponderomotive-like tidal attractions between neighboring geodesics and the shear induced attractions of the local scale factor originate from the same quantum zero-point fluctuations in the Riemann curvature.

Correspondingly, from \eqref{matter cutoff 1} and \eqref{gravity cutoff}, we have the first term $G\Lambda_{\mathrm{grav}}^4$ in \eqref{magnitude of shear}, which represents the contribution from the zero-point fluctuations of the gravitational vacuum, dominates over the second term $G^2\Lambda_{\mathrm{mat}}^6$ in \eqref{magnitude of shear}, which represents the contribution from the zero-point fluctuations of the matter fields vacuum. Then, to leading order, we have
\begin{equation}\label{sigma estimation}
\sigma^2\propto G\Lambda_{\mathrm{grav}}^4,
\end{equation}
which is the same as the leading order magnitude of $v^2$ given by \eqref{vsquare estimation}.

\section{Implications for the Cosmological Constant Problem}\label{ccp}
The conventional formulation of the cosmological constant problem does not consider the quantum fluctuations in the spacetime curvature and the extra tidal and shear attraction they generated. We show in this subsection that they are the game changers in this problem.

\subsection{Brief Review of the Conventional Formulation of the Cosmological Constant Problem}\label{review}
In this subsection, we give a brief review of the conventional formulation of the cosmological constant problem. For more extensive reviews, see, e.g., the references \cite{RevModPhys.61.1, Adler:1995vd, Carroll:2000fy, Martin:2012bt, Burgess:2013ara}.

The cosmological constant problem is essentially the problem of the gravitational consequence of the huge quantum fields vacuum energy. The conventional formulation assumes that it gravitates according to the semiclassical Einstein Field Equations
\begin{equation}\label{sefe}
R_{\mu\nu}-\frac{1}{2}Rg_{\mu\nu}=8\pi G\langle 0| T_{\mu\nu}|0\rangle,
\end{equation}
where only the expectation value of the stress energy tensor operator in the vacuum state is allowed to gravitate.

This vacuum state is supposed to satisfy the vacuum equation of state
\begin{equation}\label{veos}
\langle T_{\mu\nu}\rangle=-\langle\rho^{(\mathrm{vac})}\rangle g_{\mu\nu}.
\end{equation}
Applying the local conservation equation 
\begin{equation}\label{conservation equation}
\nabla^{\nu} T_{\mu\nu}=0
\end{equation}
to \eqref{veos} one obtains that the vacuum energy density must be a constant:
\begin{equation}
\langle\rho^{(\mathrm{vac})}\rangle=\mathrm{constant}.
\end{equation}
In this way, the gravitational effect of the quantum fields vacuum is identified with a cosmological constant.

The conventional formulation further assumes that the spacetime sourced by $\langle T_{\mu\nu}\rangle$ is homogeneous and isotropic, which can be described by the standard FLRW metric in cosmology
\begin{equation}\label{flrw}
ds^2=-dt^2+a^2(t)\left(\frac{dr^2}{1-\kappa r^2}+r^2\left(d\theta^2+\sin^2\theta d\varphi^2\right)\right),
\end{equation}
where the constant $\kappa$ represents the spatial curvature.

The evolution equation for the scale factor $a(t)$ is given by the second Friedmann equation
\begin{equation}\label{Friedmann equation}
\ddot{a}=-\frac{4\pi G}{3}\left(\rho+3p\right) a,
\end{equation}
with the initial value constraint given by the first Friedmann  equation
\begin{equation}\label{first friedmann equation}
\dot{a}^2=\frac{8\pi G\rho a^2}{3}-\kappa,
\end{equation}
where $\rho=\langle T_{00}\rangle$, $p=\frac{1}{a^2}\langle T_{ii}\rangle\, (i=1, 2, 3)$.

If the vacuum equation of state \eqref{veos} is satisfied, then the evolution equation \eqref{Friedmann equation} for the scale factor $a(t)$ becomes
\begin{equation}\label{large acceleration}
\ddot{a}=w^2 a,
\end{equation}
where
\begin{equation}\label{wsquare definition}
w^2=\frac{8\pi G}{3}\left\langle\rho^{(\mathrm{vac})}\right\rangle.
\end{equation}

The solution to \eqref{large acceleration} is
\begin{equation}\label{solution to a}
a(t)=c_1\exp\left(w t\right)+c_2\exp\left(-w t\right),
\end{equation}
where
\begin{equation}
c_1=\frac{1}{2}\left(a_0+\frac{\dot{a}_0}{w}\right),\quad c_2=\frac{1}{2}\left(a_0-\frac{\dot{a}_0}{w}\right),
\end{equation}
with $a_0=a(0)$, $\dot{a}_0=\dot{a}(0)$ satisfying the constraint
\begin{equation}
\dot{a}_0^2=w^2a_0^2-\kappa,
\end{equation}
which is obtained from the initial value constraint equation \eqref{first friedmann equation}.

Given that the observed large scale spatial curvature of the Universe is very small (flat with only 0.4 percent margin of error), it is commonly set that $\kappa=0$. In this case, the FLRW metric \eqref{flrw} can be written as
\begin{equation}\label{flat flrw}
ds^2=-dt^2+a^2(t)\left(dx^2+dy^2+dz^2\right).
\end{equation}
Then the first Friedmann equation \eqref{first friedmann equation} requires that the vacuum energy density
\begin{equation}
\left\langle\rho^{(\mathrm{vac})}\right\rangle=\frac{3H^2}{8\pi G}>0,
\end{equation}
where $H=\dot{a}/a$ is the Hubble rate.

In this scenario, the term $w^2$ in the second Friedman equation \eqref{large acceleration}, as defined by \eqref{wsquare definition}, is positive. Consequently, the relative acceleration $\ddot{a}/a>0$, indicating that the matter fields vacuum produces repulsive gravitational effect. As a result, the solution \eqref{solution to a} to \eqref{large acceleration} for the scale factor $a(t)$ reduces to
\begin{equation}\label{flat flrw solution}
a(t)=a_0 e^{H t}.
\end{equation}
The Hubble rate $H$ in the above solution \eqref{flat flrw solution} can be either positive or negative:
\begin{equation}\label{two choices}
H=\pm\sqrt{\frac{8\pi G\left\langle\rho^{(\mathrm{vac})}\right\rangle}{3}}.
\end{equation}
When $H>0$, the space is expanding, and the gravitational repulsion produced by the positive $\left\langle\rho^{(\mathrm{vac})}\right\rangle$ would speed up the expansion, causing the scale factor $a(t)$ to exponentially grow. When $H<0$, the space is contacting, and the gravitational repulsion produced by the positive $\left\langle\rho^{(\mathrm{vac})}\right\rangle$ would slow down the contraction, causing the scale factor $a(t)$ to exponentially decrease.

Observation shows that the universe is expanding, indicating that $H>0$. In 1998, it was further discovered that the expansion is not slowing down but speeding up, which implies that the universe's expansion is accelerating. The standard $\Lambda$CDM model of cosmology attributes the accelerating expansion of the universe to the gravitational repulsion resulting from a positive cosmological constant or, equivalently, from the so-called dark energy with positive energy and negative pressure. It has been hoped that the vacuum of quantum matter fields may act as this cosmological constant or dark energy responsible for the acceleration of the universe’s expansion.

Unfortunately, from \eqref{magnitude of the vacuum energy density} and \eqref{wsquare definition} we have the relative acceleration predicted from effective field theory
\begin{equation}\label{wsquare estimation}
w^2=\frac{\ddot{a}}{a}\propto G\Lambda_{\mathrm{mat}}^4,
\end{equation}
which is uber-sensitive to small scale physics. If we take $\Lambda_{\mathrm{mat}}$ to the upper limit given by \eqref{matter cutoff 1}, i.e., if we take $\Lambda_{\mathrm{mat}}$ up to the PLanck energy $E_p$, the acceleration would be greater than observation by some $120$ orders of magnitude and the whole universe would disastrously explode on the Planck time scale of $10^{-43} s$. Even if we only worry about zero-point energies in quantum chromodynamics, the discrepancy is larger than $40$ orders of magnitude \cite{RevModPhys.61.1}. One might introduce a bare cosmological constant in the Einstein equations to cancel the large acceleration produced by the enormous zero-point energies of the matter fields vacuum, but the cancellation has to be extremely precise. This is the so-called vacuum catastrophe or the cosmological constant problem.

\subsection{Equivalent Formulation of the Cosmological Constant Problem in terms of Geodesic Deviation}
To gain a deeper understanding of the underlying physical mechanism, we present an equivalent formulation of the cosmological constant problem in terms of the geodesic deviation equation \eqref{matrix form} in the free falling observer $\gamma$'s local inertial frame $\{(e_0)^a, (e_1)^a, (e_2)^a, (e_3)^a\}$, as introduced in Sec.\ref{attraction}, instead of the Friedmann equation \eqref{Friedmann equation} in the FLRW coordinate system \eqref{flrw}. 

As mentioned in the last subsection, the spacetime in the conventional formulation of the cosmological constant problem is supposed to be homogeneous and isotropic. This assumption presupposes that the Weyl curvature of the spacetime is zero. Consequently, the Weyl curvature matrix $\mathrm{C}$ defined by \eqref{Weyl curvature matrix} vanishes:
\begin{equation}\label{conventional assumption 1}
\mathrm{C}=0.
\end{equation}

In $\gamma$'s local inertial frame $\{(e_0)^a, (e_1)^a, (e_2)^a, (e_3)^a\}$, the vacuum equation of state \eqref{veos} reduces to
\begin{equation}\label{flat vacuum equation of state0}
\langle T_{\mu\nu}\rangle=-\langle\rho^{(\mathrm{vac})}\rangle \eta_{\mu\nu}.
\end{equation}
Taking average on both sides of the expression \eqref{M matrix} for the matrix $\mathrm{M}$ and using the above equation \eqref{flat vacuum equation of state0} we obtain that
\begin{equation}\label{conventional assumption 2}
\overline{\mathrm{M}}=-w^2\mathrm{I},
\end{equation}
where $w^2$ is defined by \eqref{wsquare definition}, $\mathrm{I}=\mathrm{diag}(1, 1, 1)$ is the identity matrix. 

In the conventional formulation of the cosmological constant problem, only the expectation value of the matter fields stress energy tensor is supposed to gravitate. Consequently, we have
\begin{equation}\label{conventional assumption 3}
\mathrm{M}=\overline{\mathrm{M}}.
\end{equation}
Then from \eqref{conventional assumption 1}, \eqref{conventional assumption 2} and \eqref{conventional assumption 3}, the curvature matrix $\mathrm{R}$ becomes
\begin{equation}\label{conventional R}
\mathrm{R}=\mathrm{C}+\mathrm{M}=-w^2\mathrm{I}.
\end{equation}

Substituting \eqref{conventional R} into the matrix form geodesic deviation equation \eqref{matrix form} we have
\begin{equation}\label{conventional geodesic motion}
\frac{d^2\mathbf{X}}{d\tau^2}=w^2\mathbf{X}.
\end{equation}
The above ordinary differential equations \eqref{conventional geodesic motion}, which describe the three-dimensional motion of the deviation vector $\mathbf{X}$, take exactly the same form as the ordinary different equation \eqref{large acceleration}, which describes the evolution of the scale factor $a$. The solution to \eqref{conventional geodesic motion} also takes exactly the same form as the solution \eqref{solution to a} to \eqref{large acceleration}: 
\begin{eqnarray}\label{conventional solution}
\mathbf{X}(\tau)=\mathbf{c_1}\exp\left(w\tau\right)+\mathbf{c_2}\exp\left(-w\tau\right),
\end{eqnarray}
where
\begin{eqnarray}
\mathbf{c_1}&=&\frac{1}{2}\left(\mathbf{X}(0)+\frac{1}{w}\frac{d\mathbf{X}}{d\tau}(0)\right),\\
\mathbf{c_2}&=&\frac{1}{2}\left(\mathbf{X}(0)-\frac{1}{w}\frac{d\mathbf{X}}{d\tau}(0)\right).
\end{eqnarray}

In the scenario where $\left\langle\rho^{(\mathrm{vac})}\right\rangle>0$, the first term quickly becomes dominant (except for the very special initial condition $\frac{d\mathbf{X}}{d\tau}(0)=-w\mathbf{X}(0)$, one has $\mathbf{c_1}=\mathbf{0}$) and the deviation vector $\mathbf{X}(\tau)$ goes as
\begin{equation}\label{exponential explosion}
\mathbf{X}(\tau)\sim \mathbf{c_1}\exp\left(w\tau\right).
\end{equation}
So the deviation vector $\mathbf{X}$ experiencing exponential growth, indicating that the free falling particles would quickly fly apart from each other due to the large repulsion produced by the vacuum with positive energy density and negative pressure.

\subsection{The Missing Fluctuations}
The conventional formulation of the cosmological constant problem only considered the spacetime evolution sourced by the expectation value of the matter fields stress energy tensor. It missed both the passive fluctuations of the spacetime sourced by the fluctuations in the matter fields stress energy tensor and the active fluctuations of the spacetime itself. 

More precisely, the value of the stress energy $T_{\mu\nu}$ receives contributions from the zero-point fluctuations of the quantum matter fields. By assuming the semiclassical Einstein field equations \eqref{sefe}, the conventional formulation of the cosmological constant problem only allows the expectation value $\langle T_{\mu\nu}\rangle$ produced by these zero-point fluctuations to gravitate. However, as shown in Sec.\ref{passive fluctuation}, the contributions to $T_{\mu\nu}$ from these zero-point fluctuations are not constant. The value of $T_{\mu\nu}$ itself also exhibits large zero-point fluctuation $\Delta T_{\mu\nu}$, which is on the same order of its expectation value $\langle T_{\mu\nu}\rangle$. These large passive fluctuations are missing in the conventional formulation.

Furthermore, the spacetime undergoes automatic fluctuations due to the quantum nature of gravity itself. By assuming that the spacetime is homogeneous and isotropic and using the FLRW metric \eqref{flrw}, the conventional formulation of the cosmological constant problem presupposes that the Weyl curvature is zero. However, as shown in Sec.\ref{active fluctuations}, there exists substantial zero-point fluctuations in the Weyl curvature $C^{\mu}_{\nu\lambda\rho}$ at small scales. These substantial active fluctuations are also missing in the conventional formulation.

From a physical standpoint, the large expectation value $\langle T_{\mu\nu}\rangle$ of the matter fields, along with the significant fluctuations $\Delta T_{\mu\nu}$ and $\Delta C^{\mu}_{\nu\lambda\rho}$ in both the matter fields and the Weyl curvature, are all generated from zero-point fluctuations. They originate from the same uncertainty principle. It is entirely unreasonable to solely consider the effect of the large expectation value $\langle T_{\mu\nu}\rangle$ while disregarding the effects of the substantial fluctuations $\Delta T_{\mu\nu}$ and $\Delta C^{\mu}_{\nu\lambda\rho}$. In fact, as early as 1957, the importance of these fluctuations was emphasized by Wheeler in reference \cite{Wheeler:1957mu}, stating that:

``\emph{In any case we conclude it is essential to allow for fluctuations in the metric and gravitational interactions in any proper treatment of the compensation problem---the problem of compensation of ``infinite" energies that is so central to the physics of fields and particles.}"

``\emph{This circumstance strengthens the conclusion that gravitation cannot be overlooked in any satisfactory account of the zero energy density of the vacuum.}"

Upon incorporating these fluctuations, it becomes evident that the small scale structure of spacetime deviates significantly from that described by the FLRW metric \eqref{flrw}. A reevaluation of the cosmological constant problem that encompasses these fluctuations become necessary. In the following subsections, we incorporate these fluctuations and the extra attraction they generated into the cosmological constant problem.

\subsection{Differences Made by the Fluctuations}
The conventional formulation of the cosmological constant problem is based on a crucial assumption --- the spacetime is homogeneous and isotropic. Starting from this assumption, one uses the standard FLRW metric \eqref{flrw} of cosmology to describe the spacetime dynamics given by the semiclassical Einstein equations \eqref{sefe}. While this assumption holds reasonably well at cosmological scales, it is not valid at small scales. It is important to note that the cosmological constant problem arises from quantum fluctuations occurring at small scales, specifically the Planck scale. Therefore, there is no justification for assuming that the cosmological FLRW metric is still applicable at such minute scales. In reality, both the passive fluctuations sourced by the matter fields and the active fluctuations of gravity itself lead to significant local spacetime inhomogeneity and anisotropy. Consequently, it is necessary to employ a general metric instead of the usual FLRW metric to adequately describe the substantial spacetime fluctuations.

The main points above have been previously discussed in our earlier publications, namely \cite{PhysRevD.95.103504}, \cite{PhysRevD.98.063506}, \cite{PhysRevD.102.023537} and \cite{PhysRevLett.125.051301}. In particular, in the papers \cite{PhysRevD.102.023537} and \cite{PhysRevLett.125.051301}, we generalized the FLRW metric to the general metric \eqref{general metric} to reformulate the cosmological constant problem. The importance of the small-scale inhomogeneity in the cosmological constant problem has also been studied by others, such as Carlip's works \cite{Carlip:2018zsk, Carlip_2019, Carlip:2021lau, Carlip:2021oso, Carlip:2022pyh} and Firouzjahi' works \cite{PhysRevD.106.045015, PhysRevD.106.083510, PhysRevD.108.065002}.

Next we summarize the essential conceptual and technical differences between the new formulation, based on the general metric \eqref{general metric}, and the conventional formulation, which relies on the spatially flat FLRW metric \eqref{flat flrw}.

In the homogeneous and isotropic spacetime described by the metric \eqref{flat flrw}, the physical volume $V_{\mathcal{U}}$ of a comoving region (the region with fixed spatial coordinates) $\mathcal{U}$ is straightforwardly related to the fixed comoving volume $V_{\mathcal{U}}^{(\mathrm{com})}=\int_{\mathcal{U}}d^3x$ through the simple equation:
\begin{equation}\label{simple equaiton}
V_{\mathcal{U}}(t)=a^3(t)V_{\mathcal{U}}^{(\mathrm{com})}.
\end{equation}

The Hubble rate $H$ is related to the relative rate of change of the physical volume $V_{\mathcal{U}}$ by
\begin{equation}\label{volume rate of change flrw}
H(t)=\frac{\dot{a}}{a}(t)=\frac{1}{3}\frac{\dot{V}_{\mathcal{U}}}{V_{\mathcal{U}}}.
\end{equation}

However, in the locally inhomogeneous and anisotropic spacetime described by the general metric \eqref{general metric}, a notable distinction arises. The generalized local scale factor $a(t, \mathbf{x})$ defined by \eqref{a definition}, which characterizes the local size of space at each point, becomes spatially dependent. Unlike the homogeneous and isotropic scenario, the physical volume $V_{\mathcal{U}}$ associated with the comoving region $\mathcal{U}$ is defined by \eqref{volume definition}, which can be expressed as
\begin{equation}\label{volume definition 1}
V_{\mathcal{U}}(t)=\int_{\mathcal{U}}d^3x\big|a^3(t, \mathbf{x})\big|.
\end{equation}
It is no longer related to the fixed comoving volume $V_{\mathcal{U}}^{(\mathrm{com})}$ by the simple equation given by \eqref{simple equaiton}.

Moreover, the Hubble rate, which characterizes the rate of expansion of space, also becomes spatially dependent. We can define the local Hubble rate $H^{(\mathrm{loc})}_{\mathbf{x}}(t)$ at each spatial point $\mathbf{x}$ as
\begin{equation}\label{local Hubble rate definition}
H^{(\mathrm{loc})}_{\mathbf{x}}(t)=\frac{\dot{a}}{a}(t, \mathbf{x}).
\end{equation}
At each spatial point $\mathbf{x}$, on average, the space is locally expanding if $H^{(\mathrm{loc})}_{\mathbf{x}}>0$, while the space is locally contracting if $H^{(\mathrm{loc})}_{\mathbf{x}}<0$.

Similar to \eqref{volume rate of change flrw}, we can define the global Hubble rate $H_{\mathcal{U}}^{(\mathrm{glo})}(t)$ associated with the macroscopic comoving region $\mathcal{U}$ as
\begin{equation}\label{global hubble rate definition}
H^{(\mathrm{glo})}_{\mathcal{U}}(t)=\frac{1}{3}\frac{\dot{V}_{\mathcal{U}}}{V_{\mathcal{U}}}=\frac{\int_{\mathcal{U}}d^3x\,\sqrt{h}\,H^{(\mathrm{loc})}_{\mathbf{x}}(t)}{\int_{\mathcal{U}}d^3x\,\sqrt{h}}.
\end{equation}
It characterizes the relative rate of change of the physical volume $V_{\mathcal{U}}$. If $H^{(\mathrm{glo})}_{\mathcal{U}}>0$, the comving region $\mathcal{U}$ is expanding; if $H^{(\mathrm{glo})}_{\mathcal{U}}<0$, the comving region $\mathcal{U}$ is contracting.

In the case of the homogeneous and isotropic scenario, the metric \eqref{flat flrw} has only one degree of freedom $a(t)$ for the whole space. The scale factor $a$ and the Hubble rate $\dot{a}/a$ is spatially independent and \eqref{global hubble rate definition} just reduces to \eqref{volume rate of change flrw}. Under the initial value constraint \eqref{first friedmann equation} (set $\kappa=0$ in \eqref{first friedmann equation}), there are only two distinct choices, as given by \eqref{two choices}, for the Hubble rate on the spatial slice $\Sigma_t$ defined by $t=\mathrm{constant}$. This implies that all points in space have to be simultaneously expanding or contracting at the same large constant rate.

Meanwhile, the evolution equation for the scale factor $a(t)$ is given by \eqref{large acceleration}. For convenience, we rewrite it here:
\begin{equation}\label{large acceleration 1}
\ddot{a}=w^2 a=\frac{8\pi G}{3}\left\langle\rho^{(\mathrm{vac})}\right\rangle a.
\end{equation}
It implies that the acceleration of the FLRW universe's expansion is directly proportional to the large vacuum energy density $\left\langle\rho^{(\mathrm{vac})}\right\rangle$.

However, the situation drastically changes when dealing with the locally inhomogeneous and anisotropic metric \eqref{general metric}. In this scenario, the metric \eqref{general metric} has six degrees of freedom $h_{ij}(t, \mathbf{x})$ at each point of space, implying that  the general metric \eqref{general metric} has infinitely many degrees of freedom. The initial value constraint equations are now given by \eqref{Hamiltonian constraint} and \eqref{momentum constraint}, and there exists are countless ways to select different local Hubble rates $H^{(\mathrm{loc})}_{\mathbf{x}}(t)$ at various spatial points under the constraint equations \eqref{Hamiltonian constraint} and \eqref{momentum constraint}.

Meanwhile, the evolution equation for the spatially dependent local scale factor $a(t, \mathbf{x})$ is given by \eqref{evo} instead of \eqref{large acceleration 1}, which applies solely to the spatially independent scale factor $a(t)$. For convenience, we denote the term $\frac{4\pi G}{3}\left(T_{00}+h^{ij}T_{ij}\right)$ in \eqref{evo} as $\Omega_{\mathrm{mat}}^2$:
\begin{equation}\label{Omega square definition}
\Omega_{\mathrm{mat}}^2=\frac{4\pi G}{3}\left(T_{00}+h^{ij}T_{ij}\right).
\end{equation}
It represents the contribution to the evolution of the local scale factor $a(t, \mathbf{x})$ from the matter fields. Then the evolution equation \eqref{evo} can be expressed in the following form:
\begin{equation}\label{evolution of the scale factor oscillator}
\ddot{a}=-\left(\Omega_{\mathrm{mat}}^2+\frac{2\sigma^2}{3}\right)a.
\end{equation}

We further decompose the term $-\Omega_{\mathrm{mat}}^2$ in the above equation \eqref{evolution of the scale factor oscillator} as its average $w^2$ plus fluctuation $\Delta w^2$:
\begin{equation}
-\Omega_{\mathrm{mat}}^2=w^2+\Delta w^2,
\end{equation}
where
\begin{equation}\label{wsquare definition 1}
w^2=\left\langle-\Omega_{\mathrm{mat}}^2\right\rangle=\frac{8\pi G}{3}\left\langle\rho^{(\mathrm{vac})}\right\rangle.
\end{equation}
In the above equation \eqref{wsquare definition 1}, we have used the vacuum equation of state \eqref{veos}. Then the evolution equation \eqref{evolution of the scale factor oscillator} for the local scale factor $a(t, \mathbf{x})$ can be further expressed in the following form
\begin{equation}\label{evolution equation final form}
\ddot{a}=\left(w^2+\Delta w^2-\frac{2\sigma^2}{3}\right)a.
\end{equation}

Note that the term $w^2$ defined by \eqref{wsquare definition 1}, which is calculated in the locally inhomogeneous and anisotropic spacetime described by the metric \eqref{general metric}, is exactly equal to the term $w^2$ defined by \eqref{wsquare definition}, which is calculated in the homogeneous and isotropic spacetime described by the FLRW metric \eqref{flrw}. So for convenience, we have used the same notation $w^2$ in both the evolution equation \eqref{large acceleration 1} for the spatially independent global scale factor $a(t)$ of the FLRW metric \eqref{flrw}, and the evolution equation \eqref{evolution equation final form} for the spatially dependent local scale factor $a(t, \mathbf{x})$ of the general metric \eqref{general metric}.

The differences made by the fluctuations can also be seen from the perspective of the free falling observer $\gamma$'s local inertial frame $\{(e_0)^a, (e_1)^a, (e_2)^a, (e_3)^a\}$, which is moving in the spacetime with significant curvature fluctuations. By analyzing the dynamics of the deviation vector $\mathbf{X}$ described by the geodesic deviation equation \eqref{matrix form}, the effects of these fluctuations becomes apparent. Due to these fluctuations, the deviation vector $\mathbf{X}$ does not follow the trajectory given by \eqref{conventional solution}, but rather follows an average smooth path $\overline{\mathbf{X}}$, with small oscillations $\xi_{\mathbf{X}}$ around $\overline{\mathbf{X}}$. The evolution of the smooth path $\overline{\mathbf{X}}$ is given by \eqref{average evolution equation for X with fluctuations}.

In this scenario, the average of the Weyl curvature matrix vanishes
\begin{equation}
\overline{\mathrm{C}}=0.
\end{equation}
The average of the matter fields matrix is given by
\begin{equation}
\quad \overline{\mathrm{M}}=-w^2\mathrm{I}.
\end{equation}
Consequently, the average of the curvature matrix $\overline{\mathrm{R}}$ in \eqref{average evolution equation for X with fluctuations} becomes
\begin{equation}\label{average R}
\overline{\mathrm{R}}=\overline{\mathrm{C}}+\overline{\mathrm{M}}=-w^2\mathrm{I},
\end{equation}
which coincides with the expression \eqref{conventional R} for the curvature matrix $\mathrm{R}$ if the zero-point fluctuations in the Riemann curvature are absent.

Since the quantum fluctuations in curvature exhibit no preferred directions, the extra attraction term $\overline{\mathrm{V}^2}$ in \eqref{average evolution equation for X with fluctuations} is given by the expression \eqref{V^2 matrix form}. Plugging the above expression \eqref{average R} for $\overline{\mathrm{R}}$ and the expression \eqref{V^2 matrix form} for $\overline{\mathrm{V}^2}$ into \eqref{average evolution equation for X with fluctuations}, we obtain
\begin{equation}\label{resisting equation}
\frac{d^2\overline{\mathbf{X}}}{d\tau^2}=\left(w^2-v^2\right)\overline{\mathbf{X}}.
\end{equation}
Comparing the above equation \eqref{resisting equation} to the corresponding equation \eqref{conventional geodesic motion}, where the effects of the curvature fluctuations are absent, we observe an additional tidal attraction term $v^2$ resulting from the curvature fluctuations. This mirrors the distinction between the evolution equations \eqref{evolution equation final form} and \eqref{large acceleration 1}, where an extra shear attraction term $\sigma^2$ is present in \eqref{evolution equation final form}.

\subsection{The Role of the Extra Attraction}\label{avoidance of vacuum catastrophe}
The evolution equation \eqref{evolution equation final form} unveils the fundamental distinction in spacetime dynamics caused by the fluctuations when compared to the dynamics without the fluctuations described by equation \eqref{large acceleration 1}. More specifically, in \eqref{large acceleration 1} the relative acceleration $\ddot{a}/a$ is equal to $w^2$, while in \eqref{evolution equation final form} the relative acceleration $\ddot{a}/a$ is equal to $w^2+\Delta w^2-2\sigma^2/3$.

The term $w^2$ in \eqref{evolution equation final form} represents the contribution from the large expectation value of the energy density $\left\langle\rho^{(\mathrm{vac})}\right\rangle$ generated by the zero-point fluctuations of the matter fields vacuum. For the situation that $\left\langle\rho^{(\mathrm{vac})}\right\rangle>0$, we have $w^2>0$, resulting in strong gravitational repulsion. The term $\Delta w^2$ in \eqref{evolution equation final form} represents the contribution from the large zero-point fluctuations in the value of the matter fields vacuum stress energy. By definition, the expectation value of $\Delta w^2$ is zero. The term $-2\sigma^2/3$ in \eqref{evolution equation final form} represents the contribution from the shear induced by the zero-point fluctuations in the Riemann curvature through \eqref{shear evolution}. This term always give rise to gravitational attraction.

From \eqref{wsquare estimation} we have $w^2$ goes as $G\Lambda_{\mathrm{mat}}^4$ while from \eqref{sigma estimation} we have $\sigma^2$ goes as $G\Lambda_{\mathrm{grav}}^4$. They both go as the fourth power of the cutoff. Then from \eqref{matter cutoff} (or \eqref{matter cutoff 1}) and \eqref{gravity cutoff}, we have $\Lambda_{\mathrm{mat}}\ll\Lambda_{\mathrm{grav}}$ so that $w^2\ll \sigma^2$ and thus 
\begin{equation}\label{dominance}
w^2+\Delta w^2-2\sigma^2/3 \ll 0,
\end{equation}
implying that the gravitational attraction generated by the shear dominates over the repulsion produced by the matter fields vacuum.

Consequently, the dynamics of the spatially dependent local scale factor $a(t, \mathbf{x})$ described by \eqref{evolution equation final form} is entirely opposite to the dynamics of the spatially independent scale factor $a(t)$ described by \eqref{large acceleration 1}. The positive value of $w^2$ in \eqref{large acceleration 1} yields strong gravitational repulsion, resulting in exponential evolution of $a(t)$. Conversely, the negative value of $w^2+\Delta w^2-2\sigma^2/3$ in \eqref{evolution equation final form} generates strong gravitational attraction, leading to oscillatory evolution of $a(t, \mathbf{x})$.

More specifically, for the two choices of the initial value of $H$ given by \eqref{two choices}, the solution to \eqref{large acceleration 1} is
\begin{equation}\label{flrw solution for a}
a(t)=a_0 \exp\left(\pm\left(\frac{8\pi G}{3}\left\langle\rho^{(\mathrm{vac})}\right\rangle\right)^{\frac{1}{2}} t\right).
\end{equation}
And, from \eqref{simple equaiton} and the above solution \eqref{flrw solution for a} for $a(t)$, we have the physical volume
\begin{eqnarray}\label{exponential physical volume}
V_{\mathcal{U}}(t)=a_0^3\exp\left(\pm\left(24\pi G\left\langle\rho^{(\mathrm{vac})}\right\rangle\right)^{\frac{1}{2}}t\right)V_{\mathcal{U}}^{(\mathrm{com})}.
\end{eqnarray}

In contrast, since $w^2+\Delta w^2-2\sigma^2/3$ takes negative values, \eqref{evolution equation final form} describes a harmonic oscillator with a varying frequency. The most basic behavior of a harmonic oscillator is that it oscillates back and forth around its equilibrium point, which implies that the local Hubble rates $H^{(\mathrm{loc})}_{\mathbf{x}}$ defined by \eqref{local Hubble rate definition} (quasi-)periodically change signs over time. Meanwhile, due the wild fluctuations of the spacetime, the phases of the oscillations of $a(t, \mathbf{x})$ at neighboring spatial points $\mathbf{x}$ must be distinct, and the local Hubble rates $H^{(\mathrm{loc})}_{\mathbf{x}}$ also change signs over spatial directions.

Physically, these fluctuating features of $H^{(\mathrm{loc})}_{\mathbf{x}}$ imply that, at any instant of time, if the space is expanding in a small region, it has to be contracting in neighboring regions; and at any spatial point, if the space is expanding now, it has to be contracting later.

These features result in substantial cancellations when calculating the global Hubble rate $H_{\mathcal{U}}^{(\mathrm{glo})}$ associated with the macroscopic volume $\mathcal{U}$. Although the absolute value of the local Hubble rate $\big|H^{(\mathrm{loc})}_{\mathbf{x}}\big|$ at each individual point has to be enormous to satisfy the constraint equations \eqref{Hamiltonian constraint} and \eqref{momentum constraint}, the overall net Hubble rate can be small . In other words, although the instantaneous rates of expansion or contraction at a fixed spatial point can be significant, their effects can cancel out in such a way that the macroscopic scale physical volume $V_{\mathcal{U}}$ defined by \eqref{volume definition 1} associated with the comving region $\mathcal{U}$ does not grow $10^{120}$ times larger than what is observed.

More precisely, the solution to \eqref{evolution equation final form} for the local scale factor $a(t, \mathbf{x})$ would take the form
\begin{equation}\label{oscillation form}
a(t, \mathbf{x})=A(t)P(t, \mathbf{x}),
\end{equation}
where $A(t)>0$ represents the oscillation amplitude of $a(t, \mathbf{x})$, and $P(t, \mathbf{x})$ is a function that oscillates around $0$ with unit amplitude\footnote{One might be concerned that the local scale factor $a(t, \mathbf{x})$ crosses $0$ and take both positive and negative values. However, this oscillatory behavior of $a(t, \mathbf{x})$ is just a natural consequence of the evolution equation \eqref{evolution equation final form} with negative $w^2+\Delta w^2-2\sigma^2/3$. It is important to remember that $a(t, \mathbf{x})$ is defined by $h=a^6$, ensuring that mathematically, it is legitimate for $a(t, \mathbf{x})$ to take negative values. This is because the determinant $h$ of the metric and the physical volume element $\sqrt{h}d^3x$ are always non-negative. Furthermore, for the usual FLRW metric \eqref{flat flrw}, it is also legitimate for the scale factor $a(t)$ to take negative values, since the metric is not $a(t)$ but $a^2(t)$, which is always non-negative. If one transforms $a(t)$ to $-a(t)$, the FLRW metric \eqref{flat flrw} describes exactly the same spacetime.}.

For a macroscopic region $\mathcal{U}$ that is much larger than the scale of the curvature fluctuations, the function $P(t, \mathbf{x})$, which has a constant oscillation amplitude, possesses the property
\begin{equation}
\int_{\mathcal{U}} d^3x \big|P^3(t, \mathbf{x})\big|=\mathrm{constant}.
\end{equation}
Then the physical volume $V_{\mathcal{U}}$ defined by \eqref{volume definition 1} would be proportional to the third power of the oscillation amplitude $A(t)$ of the local scale factor:
\begin{equation}
V_{\mathcal{U}}(t)=A^3(t)\int_{\mathcal{U}} d^3x \big|P^3(t, \mathbf{x})\big|\propto A^3(t),
\end{equation}
and the observed macroscopic global Hubble rate, as defined by \eqref{global hubble rate definition}, would be equal to the relative rate of change of $A(t)$:
\begin{equation}\label{Hubble rate and amplitude}
H_{\mathcal{U}}^{(\mathrm{glo})}(t)=\frac{1}{3}\frac{\dot{V}_{\mathcal{U}}}{V_{\mathcal{U}}}=\frac{\dot{A}}{A}(t).
\end{equation}

It is worth noting that the macroscopic region $\mathcal{U}$ does not need to be exceedingly large; even a cubic centimeter contains about $10^{100}$ Planck-size fluctuating regions. For a sufficiently large $\mathcal{U}$, the global Hubble rate $H_{\mathcal{U}}^{(\mathrm{glo})}$ becomes independent of the size of $\mathcal{U}$, allowing us to define the observed large scale Hubble rate $H^{\mathrm{(obs)}}$ as
\begin{equation}\label{observed Hubble rate}
H^{\mathrm{(obs)}}=H^{(\mathrm{glo})}_{\mathcal{U}},
\end{equation}
when $\mathcal{U}$ is sufficiently large.

If we disregard the weak parametric resonance effect resulting from the zero-point fluctuations, which will be discussed in the next subsection, the oscillation amplitude $A$ of the local scale factor $a(t, \mathbf{x})$ would stay constant and the time derivative $\dot{A}$ in \eqref{Hubble rate and amplitude} equals zero. As a result, we have the observed large scale Hubble rate is zero and the macroscopic physical volume does not change with time:
\begin{eqnarray}
&&H^{\mathrm{(obs)}}=H^{(\mathrm{glo})}_{\mathcal{U}}=0,\\
&&V_{\mathcal{U}}(t)=\mathrm{constant}.
\end{eqnarray}

The above results exhibit a stark contrast to the violent exponential expansion or contraction of the whole space predicted by the conventional formulation of the cosmological constant problem (as depicted in equation \eqref{exponential physical volume}), which overlooks the extra attraction generated by the quantum zero-point fluctuations of the gravitational vacuum.

This vital role played by the fluctuation induced extra attractions can also be seen from the dynamics of the average path $\overline{\mathbf{X}}$ of the Jacobi filed $\mathbf{X}$ described by the evolution equation \eqref{resisting equation}. 

From \eqref{wsquare estimation} we have $w^2$ goes as $G\Lambda_{\mathrm{mat}}^4$ while from \eqref{vsquare estimation} we have $v^2$ goes as $G\Lambda_{\mathrm{grav}}^4$. They both go as the fourth power of the cutoff. Then from \eqref{matter cutoff} (or \eqref{matter cutoff 1}) and \eqref{gravity cutoff}, we have $\Lambda_{\mathrm{mat}}\ll\Lambda_{\mathrm{grav}}$ so that $w^2\ll v^2$ and thus
\begin{equation}
w^2-v^2\ll 0,
\end{equation}
implying that the gravitational tidal attraction generated by the curvature fluctuations dominates over the repulsion produced by the matter fields vacuum.

Therefore, the dynamics of $\overline{\mathbf{X}}$ described by \eqref{resisting equation} with the inclusion of the curvature fluctuations exhibits a stark contrast to the dynamics of $\mathbf{X}$ described by \eqref{conventional geodesic motion} in the absence of fluctuations. The positive $w^2$ in \eqref{conventional geodesic motion} gives rise to strong gravitational repulsion, resulting in the exponential evolution of $\mathbf{X}$ given by \eqref{exponential explosion}. Conversely, the negative $w^2-v^2$ in \eqref{resisting equation} generates strong gravitational attraction, leading to circular motion of $\overline{\mathbf{X}}$ around $\gamma$, preventing the free falling test particles from dispersing. 

In this way, the effects of the enormous gravitational repulsion produced by the matter fields vacuum are completely screened by the even more substantial attraction generated by the zero-point fluctuations in the spacetime curvature. As a result, the predicted catastrophic explosion of the universe caused by the huge positive energy and negative pressure of the matter fields vacuum is avoided.

It is important to emphasize that one of our key points is that there is no need to reduce the value of the vacuum energy density, which is inherently generated by the zero-point fluctuations in the matter fields associated with the Planck scale. The vacuum energy density remains as large as predicted by effective field theory. Despite its enormous magnitude, it does not actually result in a disastrous explosion of the universe, as the strong repulsion it generates are suppressed by the substantial attraction produced by the zero-point fluctuations of the gravitational field vacuum.

Note that this suppression is not precise in the sense that the strength of the extra attraction generated by the curvature fluctuations is exactly equal to the strength of the repulsion produced by the matter fields vacuum. In the literature, numerous attempts to solve the cosmological constant problem by precisely canceling the large energy density have been proven to be extremely challenging and inevitably require fine-tuning of theory parameters.

In our work, including this paper and the preceding ones (\cite{PhysRevD.95.103504, PhysRevD.98.063506, PhysRevD.102.023537, PhysRevLett.125.051301}), we do not rely on exact cancellation of the large energy density because the small scale structure of the fluctuating spacetime, described by the locally inhomogeneous and anisotropic metric \eqref{general metric}, differs significantly from the conventional homogeneous and isotropic spacetime described by the FLRW metric \eqref{flrw}. Rather than the space expanding or contracting as a whole, as described by the FLRW metric \eqref{flrw}, locally, at each point, the space oscillates alternatively between expansion and contraction, with varying phases of oscillation at neighboring points. These local expansions and contractions are violent at small scales. However, upon averaging over larger macroscopic scales, their effects nearly cancel out. In this way, the gravitational effects of the substantial matter fields vacuum energy density are naturally concealed within the violent small scale spacetime fluctuations.

\subsection{The Acceleration of the Universe's Expansion}
The dynamics of the local scale factor $a(t, \mathbf{x})$ is described by \eqref{evolution of the scale factor oscillator}. For the reader's convenience, we rewrite it here in the following form:
\begin{equation}\label{local scale factor evolution equation}
\ddot{a}+\left(\Omega_{\mathrm{mat}}^2+\frac{2\sigma^2}{3}\right)a=0.
\end{equation}
The evolution equation for $\sigma^2$ can be obtained from \eqref{shear evolution}. The result is
\begin{equation}
\left(\sigma^2\right)^{\cdot}=-6\frac{\dot{a}}{a}\sigma^2-R_{ij}^{(3)}\sigma^{ij}+8\pi G T_{ij}\sigma^{ij}.
\end{equation}

Note that we have $\sigma^2\,(\sim G\Lambda_{\mathrm{grav}}^4)\gg\Omega_{\mathrm{mat}}^2\,(\sim G\Lambda_{\mathrm{mat}}^4)$, i.e., the contribution to the evolution of $a(t, \mathbf{x})$ from the gravitational field vacuum dominates over the matter fields vacuum. So the term $\Omega_{\mathrm{mat}}^2$ in \eqref{local scale factor evolution equation} may be regarded as an external small perturbation to the the evolution of $a(t, \mathrm{x})$.

As discussed in the last subsection, the local scale factor $a(t, \mathbf{x})$ vibrates back and forth around $0$. The time scale of this vibration is characterized by
\begin{equation}
T_1 \sim  \frac{1}{\sigma} \sim \frac{1}{\sqrt{G}\Lambda_{\mathrm{grav}}^2}.
\end{equation}

The external small perturbation term $\Omega_{\mathrm{mat}}^2$ also vibrates (fluctuates). The time scale of this vibration is 
\begin{equation}
T_2 \sim 1/\Lambda_{\mathrm{mat}}.
\end{equation}

It is well known that when a dynamical system consists of two vibrating components, the vibration of one of the components may influence the other, leading to the occurrence of parametric resonance \cite{Kurmann17}. Similar to our previous papers \cite{PhysRevD.95.103504, PhysRevD.102.023537}, the vibration of the perturbation term $\Omega_{\mathrm{mat}}^2$ in this system would influence the vibration of the local scale factor $a(t, \mathbf{x})$, leading to the parametric amplification of the vibration amplitude $A$ (defined by \eqref{oscillation form}) of the local scale factor $a(t, \mathbf{x})$:
\begin{equation}\label{amplitude growth}
A(t)=A(0)e^{Ht}.
\end{equation}
Here the exponent $H$ characterizes the strength of the parametric amplification. By employing a similar analysis to that conducted in \cite{PhysRevD.95.103504, PhysRevD.98.063506, PhysRevD.102.023537, PhysRevLett.125.051301}, we obtain that the magnitude of $H$ depends on the relative magnitude between the time scales $T_1$ and $T_2$:
\begin{eqnarray}
H &=& \alpha\Lambda_{\mathrm{mat}}\exp\left(-\beta T_2/T_1\right)\nonumber\\
&=&\alpha\Lambda_{\mathrm{mat}}\exp\left(-\beta \sqrt{G}\Lambda_{\mathrm{grav}}^2/\Lambda_{\mathrm{mat}}\right)\label{new hubble expansion rate}\\
&=&\alpha\Lambda_{\mathrm{mat}}\exp\left(-\beta \left(\frac{\Lambda_{\mathrm{grav}}}{E_p}\right)\left(\frac{\Lambda_{\mathrm{grav}}}{\Lambda_{\mathrm{mat}}}\right)\right),\nonumber
\end{eqnarray}
where $\alpha, \beta>0$ are two dimensionless parameters whose exact value depend on the detailed property of the quantum zero-point fluctuations. 

Inserting \eqref{amplitude growth} into \eqref{oscillation form}, we have the evolution of the local scale factor takes the form
\begin{equation}
a(t, \mathbf{x})=A(0)e^{Ht}P(t, \mathbf{x}),
\end{equation}
with the exponent $H$ given by \eqref{new hubble expansion rate}. Then the physical volume $V_{\mathcal{U}}$ associated with the macroscopic region $\mathcal{U}$, which is defined by \eqref{volume definition 1}, becomes
\begin{equation}
V_{\mathcal{U}}(t)=V_{\mathcal{U}}(0)e^{3Ht},
\end{equation}
where
\begin{equation}
V_{\mathcal{U}}(0)=A^3(0)\int_{\mathcal{U}}d^3x\big|P^3(t, \mathbf{x})\big|=\mathrm{constant}.
\end{equation}

Thus, after averaging the vibrations of the local scale factor over macroscopic scales, we obtain an exponentially expanding universe with the expansion rate given by \eqref{new hubble expansion rate}. More precisely, inserting \eqref{amplitude growth} into \eqref{Hubble rate and amplitude} and using \eqref{observed Hubble rate}, we have the observed large scale Hubble expansion rate
\begin{equation}
H^{\mathrm{(obs)}}=H_{\mathcal{U}}^{(\mathrm{glo})}=\frac{1}{3}\frac{\dot{V}_{\mathcal{U}}}{V_{\mathcal{U}}}=H.
\end{equation}

The value of $H$ given by \eqref{new hubble expansion rate} is in contrast to the conventional value given by
\begin{eqnarray}
H&=&\sqrt{\frac{8\pi G\left\langle\rho^{(\mathrm{vac})}\right\rangle}{3}}\sim \sqrt{G}\Lambda_{\mathrm{mat}}^2 \nonumber\\
&\to & +\infty, \quad\text{as}\quad\Lambda_{\mathrm{mat}}\to +\infty,
\end{eqnarray}
which is based on the homogeneous and isotropic FLRW metric \eqref{flat flrw}.



The new result \eqref{new hubble expansion rate} establishes a connection between physics at the smallest quantum gravity scale and the largest cosmological scale. It naturally gives rise to the slowly accelerating expansion of the universe. Physically, this happens because the small scale structure of the spacetime described by the general metric \eqref{general metric} is drastically different from the conventional FLRW metric \eqref{flat flrw} and the strength of the parametric resonance effect given by \eqref{new hubble expansion rate} which leads to the accelerated expansion is very weak. This result suggests that there is no necessity to introduce the bare cosmological constant, which requires fine-tuning to an accuracy of $10^{-120}$, or other forms of dark energy, which are required to have peculiar negative pressure, to explain the observed accelerating expansion of the universe.

\section{Discussions}\label{small scale structure}
In this section we discuss some particular issues of the cosmological constant problem and the differences between the work presented in this paper and the previous ones.

\subsection{The Sign Ambiguities}
Our discussions in the last section.\ref{ccp} have adopted two conventional assumptions of the cosmological constant problem: (i) the expectation value of the vacuum energy density $\left\langle\rho^{(\mathrm{vac})}\right\rangle$ is positive; (ii) the zero-point fluctuations of the matter fields vacuum satisfy the vacuum equation of state \eqref{veos}. In other words, the gravitational effect of the expectation value of the matter fields stress energy generated by the zero-point fluctuations is supposed to  be identical to a large positive cosmological constant.

However, there are ambiguities in the above two conventional assumptions. One of the main distinctions between gravity and other fundamental forces lies in the fact that gravity couples to everything. According to the equivalence principle of general relativity, all (known and unknown) fundamental fields gravitate and contribute to $\left\langle\rho^{(\mathrm{vac})}\right\rangle$. In particular, Boson fields give positive contributions to $\left\langle\rho^{(\mathrm{vac})}\right\rangle$, while Fermi fields give negative contributions to $\left\langle\rho^{(\mathrm{vac})}\right\rangle$, so the sign of $\left\langle\rho^{(\mathrm{vac})}\right\rangle$ depends on the relative magnitudes of these contributions \cite{Martin:2012bt}. Although the known fundamental fields have been described by the standard model of particle physics, we are uncertain about the existence of other fundamental fields beyond this model. Thus, the sign of $\left\langle\rho^{(\mathrm{vac})}\right\rangle$ is undetermined.

Furthermore, each contribution to $\left\langle\rho^{(\mathrm{vac})}\right\rangle$ is formally divergent, regularization is needed to obtain a finite value for $\left\langle\rho^{(\mathrm{vac})}\right\rangle$. Whether the zero-point fluctuations of the matter fields vacuum satisfy the vacuum equation of state \eqref{veos} hinges on the Lorentz invariance of the regularization scheme. However, the question of whether these zero-point fluctuations of matter fields maintain Lorentz invariance, especially in the wildly fluctuating spacetime, remains unclear \cite{PhysRevD.97.068302}. Consequently, whether the vacuum equation of state \eqref{veos} should be satisfied is an open question.

If one or both of the above two conventional assumptions (i) and (ii) are violated, the gravitational effect of the expectation value of the matter fields vacuum is no longer identical to a positive cosmological constant. In particular, if assumption (ii) is not true, it does not behave like a cosmological constant, and, strictly speaking, the problem it gives rise to should not be referred to as the cosmological constant problem. However, the zero-point fluctuation induced expectation value of the matter fields stress energy remains large and it may still cause problems.

These issues are rarely explored in the literature. In the following discussions, we delve into these matters and investigate whether our proposal for addressing the cosmological constant problem remain effective if one or both of the above two conventional assumptions are violated.

\subsubsection{The Sign of $\langle T_{00}\rangle$ in the Standard Model}
To illustrate the sign ambiguities mentioned above, let us consider the simplest case of free fields in Minkowski spacetime. Direct calculations yield the following results:
\begin{eqnarray}
\langle T_{00}\rangle_{\mathrm{Minkowski}}&=&\pm\frac{n}{(2\pi)^3}\int d^3k \frac{\omega}{2}\label{vacuum energy},\\
\langle T_{ii}\rangle_{\mathrm{Minkowski}}&=&\pm\frac{n}{(2\pi)^3}\frac{1}{3}\int d^3k\frac{\mathbf{k}^2}{2\omega},\label{vacuum pressure}\\
\langle T_{0i}\rangle_{\mathrm{Minkowski}}&=&\langle T_{ij}\rangle_{\mathrm{Minkowski}}=0,\, i\neq j,
\end{eqnarray}
where $i, j=1, 2, 3$, $\omega=\sqrt{\mathbf{k}^2+m^2}$ is the dispersion relation, the positive sign ``$+$" corresponds to Boson fields while the minus sign ``$-$" corresponds to Fermi fields, the degeneracy factor $n$ includes contributions from spin, such that $n=2s+1$ for massive particles, $n=2$ for massless particles, an additional factor of $2$ if particle and antiparticle are distinct and an additional factor of $3$ due to color \cite{Martin:2012bt, Visser:2016mtr}.

Next, let us count the contributions to the degeneracy factor $n$ in the expression \eqref{vacuum energy} for $\langle T_{00}\rangle$ from the fundamental fields in the standard model. For the massless photon with spin $s=1$, we have
\begin{equation}
n_{\mathrm{photon}}=2\times 1=2.
\end{equation}
For the massive W and Z Bosons with spin $s=1$, we have
\begin{equation}
n_{\mathrm{W^{\pm},Z}}=(2\times 1+1)\times 3=9.
\end{equation}
For the eight types of massless gluons with spin $s=1$, we have
\begin{equation}
n_{\mathrm{gluons}}=2\times 8=16.
\end{equation}
For the massive Higgs Boson with spin $s=0$, we have
\begin{equation}
n_{\mathrm{Higgs}}=2\times 0+1=1.
\end{equation}
For the six massive quark flavors with spin $s=\frac{1}{2}$, distinct aniquarks and three colors, we have
\begin{equation}
n_{\mathrm{quarks}}=(2\times\frac{1}{2}+1)\times 2\times 3\times 6=72.
\end{equation}
For the six massive leptons with spin $s=\frac{1}{2}$ and distinct antileptons, we have
\begin{equation}
n_{\mathrm{leptons}}=(2\times\frac{1}{2}+1)\times 2\times 6=24.
\end{equation}
Then we have the degeneracy factor for the bosons
\begin{eqnarray}
n_{\mathrm{bosons}}&=&n_{\mathrm{photon}}+n_{\mathrm{W^{\pm},Z}}+n_{\mathrm{gluons}}+n_{\mathrm{Higgs}}\nonumber\\
&=&2+9+16+1\nonumber\\
&=&28,
\end{eqnarray}
and, the degeneracy factor for the fermions
\begin{eqnarray}
n_{\mathrm{fermions}}&=&n_{\mathrm{quarks}}+n_{\mathrm{leptons}}\nonumber\\
&=&72+24\nonumber\\
&=&96.
\end{eqnarray}

So we have $n_{\mathrm{fermions}}>n_{\mathrm{bosons}}$, which means that there are more Fermi fields degree of freedom than the Boson fields. Consequently, for the fundamental fields in the standard model, the contributions to $\langle T_{00}\rangle$ from the Fermi fields are larger than from the Boson fields and we have $\langle T_{00}\rangle<0$.

\subsubsection{The Sign of $\left\langle\Omega_{\mathrm{mat}}^2\right\rangle$}
The term $\Omega_{\mathrm{mat}}^2$ defined by \eqref{Omega square definition} represents the contributions from the matter fields to the evolution of the local scale factor $a(t, \mathbf{x})$ given by \eqref{evolution of the scale factor oscillator}. From \eqref{vacuum energy} and \eqref{vacuum pressure} we have
\begin{equation}\label{Minkowski Omega}
\left\langle\Omega_{\mathrm{mat}}^2\right\rangle_{\mathrm{Minkowski}}=\pm\frac{4\pi G}{3}\frac{n}{(2\pi)^3}\int d^3k \left(\frac{\omega}{2}+\frac{\mathbf{k}^2}{2\omega}\right).
\end{equation}
The integrand $\frac{\omega}{2}+\frac{\mathbf{k}^2}{2\omega}$ in the above expression \eqref{Minkowski Omega} is positive, so we should have $\left\langle\Omega_{\mathrm{mat}}^2\right\rangle_{\mathrm{Minkowski}}>0$ for Boson fields and $\left\langle\Omega_{\mathrm{mat}}^2\right\rangle_{\mathrm{Minkowski}}<0$ for Fermi fields.

However, this outcome contradicts the supposed vacuum equation of state \eqref{veos}. If \eqref{veos} holds true, then we have $\left\langle T_{ii}\right\rangle_{\mathrm{Minkowski}}=-\left\langle T_{00}\right\rangle_{\mathrm{Minkowski}}$, leading to
\begin{eqnarray}
\left\langle\Omega_{\mathrm{mat}}^2\right\rangle_{\mathrm{Minkowski}}=-\frac{8\pi G}{3}\left\langle T_{00}\right\rangle_{\mathrm{Minkowski}}\nonumber\\
=\mp\frac{8\pi G}{3}\frac{n}{(2\pi)^3}\int d^3k\frac{\omega}{2}.
\end{eqnarray}
So we should have $\left\langle\Omega_{\mathrm{mat}}^2\right\rangle_{\mathrm{Minkowski}}<0$ for Boson fields and $\left\langle\Omega_{\mathrm{mat}}^2\right\rangle_{\mathrm{Minkowski}}>0$ for Fermi fields, which has an opposite sign to \eqref{Minkowski Omega}.

This contradiction originates from the ambiguities in evaluating the divergent integrals \eqref{vacuum energy} and \eqref{vacuum pressure}. These divergences are the root causes of the cosmological constant problem. If we evaluate \eqref{vacuum energy} and \eqref{vacuum pressure} by imposing a large momentum cutoff $|\mathbf{k}|=\Lambda_{\mathrm{mat}}\gg m$, then we obtain the following result
\begin{equation}\label{radiation state}
\langle T_{\mu\nu}\rangle_{\mathrm{Minkowski}}= \pm\mathrm{diag}\left(1, \frac{1}{3}, \frac{1}{3}, \frac{1}{3}\right)\times\frac{n}{16\pi^2}\Lambda_{\mathrm{mat}}^4,
\end{equation}
which does not obey the supposed vacuum equation of state \eqref{veos}. It behaves more like a radiation field.

It is argued that this violation of the vacuum equation of state originates from the use of the noncovariant cutoff $\Lambda_{\mathrm{mat}}$. If one uses regularization methods which respect the Lorentz invariance, such as dimensional regularization, to evaluate \eqref{vacuum energy} and \eqref{vacuum pressure}, and then follows a $\overline{\mathrm{MS}}$ renormalzation scheme to subtract the divergent part, one obtains $T_{\mu\nu}\propto m^4\ln(m^2/\mu^2)g_{\mu\nu}$, with $\mu$ the renormalization scale \cite{Martin:2012bt}. This result does obey the supposed vacuum equation of state \eqref{veos}, but it predicts that massless fields have no contribution to $\langle T_{\mu\nu}\rangle$, which should, in principle, possess nonzero zero-point energies.

Thus far, we have demonstrated that even in the simplest case of Minkowski spacetime, sign ambiguities persist, let alone in the case of more intricate fluctuating curved spacetime. It is improbable that these sign ambiguities will disappear in the case of curved spacetime.

\subsubsection{Lorentz Invariance of the Zero-Point Fluctuations?}
The vacuum equation of state \eqref{veos} is obtained from the Lorentz invariance argument. More precisely, it is argued that the vacuum is Lorentz invariant and thus every inertial observer sees the same vacuum. All physical properties extracted from this vacuum state, such as the expectation value of the stress energy tensor, should also remain invariant under Lorentz transformations. Since in Minkowski spacetime, $\eta_{\mu\nu}$ is the only Lorentz invariant $(0,2)$ tensor up to a constant, the expectation value of the stress energy tensor must be proportional to $\eta_{\mu\nu}$:
\begin{equation}\label{flat vacuum equation of state}
\langle T_{\mu\nu}\rangle_{\mathrm{Minkowski}}=-\langle\rho^{(\mathrm{vac})}\rangle \eta_{\mu\nu}.
\end{equation}
The above equation is then straightforwardly generalized, by the principle of general covariance, to the vacuum equation of state \eqref{veos} in curved spacetime. 

However, the validity of the above reasoning is questionable.

First, it is well known that there is no well defined vacuum state in general curved spacetime due to the absence of consensus among free falling observers regarding the choice of vacuum. It is even unclear that whether there exists a Lorentz invariant state in the wildly fluctuating spacetime considered in this paper. In the absence of a well-defined vacuum concept (which may not exist at all), one can not provide a precise meaning of $\langle 0|T_{\mu\nu}|0\rangle$, let alone assert that this quantity is Lorentz invariant and satisfies the vacuum equation of state \eqref{veos}.

Second, even in the case of Minkowski spacetime where the vacuum state is well defined, it is dubious to assume that the fluctuation induced zero-point energy is Lorentz invariant.

To illustrate this point, let us revisit the definition of the vacuum state. For simplicity, consider a quantized free massless scalar field $\phi$ in $1+1$ dimensional spacetime. We express $\phi$ in a Cartesian coordinate system $(t, x)$ as follows:
\begin{equation}\label{1+1 field}
\phi(t, x)=\int_{-\infty}^{+\infty}\frac{dk}{\sqrt{4\pi|k|}}\left(a_ke^{-i(|k|t-kx)}+a_k^{\dag}e^{i(|k|t-kx)}\right).
\end{equation}
The vacuum state is defined as
\begin{equation}
a_k|0\rangle=0,
\end{equation}
for any $k\in (-\infty, +\infty)$.

The expression
\begin{equation}\label{dphi}
\phi^{(k,\,k+dk)}=\frac{dk}{\sqrt{4\pi|k|}}\left(a_ke^{-i(|k|t-kx)}+a_k^{\dag}e^{i(|k|t-kx)}\right)
\end{equation}
inside the integral in \eqref{1+1 field} represents the contribution to $\phi$ from the field modes with wave numbers between $k$ and $k+dk$. The expectation value of the zero-point energy produced by the zero-point fluctuations of these modes is
\begin{eqnarray}
E_0^{(k, \,k+dk)}&=&\frac{1}{2}\left\langle\left(\frac{\partial\phi}{\partial t}\right)^2+\left(\frac{\partial\phi}{\partial x}\right)^2\right\rangle dk\nonumber\\
&=&\frac{|k|}{4\pi}dk.\label{E0k}
\end{eqnarray}
The total zero-point energy is defined as the integration of $E_0^{(k, \,k+dk)}$ over all $k$ from $-\infty$ to $+\infty$,
\begin{equation}\label{totoal zero}
E_0=\int_{-\infty}^{+\infty}E_0^{(k, \,k+dk)}=\int_{-\infty}^{+\infty}\frac{|k|}{4\pi}dk.
\end{equation}

If we perform a Lorentz boost and transform the field mode $e^{\pm i(|k|t-kx)}$ to the Cartesian coordinate system $(t', x')$ associated with an observer moving along the trajectory $x=vt$, the wave number $k$ in the $(t, x)$ system will be Doppler shifted to the wave number $k'$ in the $(t', x')$ system according to
\begin{equation}\label{Doppler shift}
k'=
  \begin{cases}
    \left(\frac{1+v}{1-v}\right)^{\frac{1}{2}}k,       & \quad \text{if } k<0,\\
    \left(\frac{1-v}{1+v}\right)^{\frac{1}{2}}k,  & \quad \text{if } k>0.
  \end{cases}
\end{equation}
Meanwhile, the operator coefficient $a_{k}$ associated with the $(t, x)$ system will be transformed to $b_{k'}$ in the $(t', x')$ system according to
\begin{equation}
b_{k'}=
  \begin{cases}
    \left(\frac{1-v}{1+v}\right)^{\frac{1}{4}}a_k,  & \quad \text{if } k<0,\\
    \left(\frac{1+v}{1-v}\right)^{\frac{1}{4}}a_k,  & \quad \text{if } k>0.
  \end{cases}
\end{equation}

Correspondingly, the expression \eqref{dphi} for $\phi^{(k, \,k+dk)}$, which represents the contribution to $\phi$ from the field modes with wave numbers between $k$ and $k+dk$ in the $(t, x)$ system, is then transformed to
\begin{eqnarray}
&&\phi^{(k', \,k'+dk')}\\
=&&\frac{dk'}{\sqrt{4\pi|k'|}}\left(b_{k'}e^{-i(|k'|t'-k'x')}+b_{k'}^{\dag}e^{i(|k'|t'-k'x')}\right),\nonumber
\end{eqnarray}
which represents the contribution to $\phi$ from the field modes of wave numbers between $k'$ and $k'+dk'$ in the $(t', x')$ system. Integrate these contributions over all $k'$, we obtain the expression for $\phi$ in the $(t', x')$ system:
\begin{eqnarray}\label{1+1' field}
&&\phi(t', x')\\
=&&\int_{-\infty}^{+\infty}\frac{dk'}{\sqrt{4\pi|k'|}}\left(b_{k'}e^{-i(|k'|t'-k'x')}+b_{k'}^{\dag}e^{i(|k'|t'-k'x')}\right).\nonumber
\end{eqnarray}

Correspondingly, $E_0^{(k, \,k+dk)}$ defined by \eqref{E0k} in the $(t, x)$ system is transformed to $E_0^{'(k', \,k'+dk')}$ in the $(t', x')$ system:
\begin{eqnarray}
E_0^{'(k', \,k'+dk')}&=&\frac{1}{2}\left\langle\left(\frac{\partial\phi}{\partial t'}\right)^2+\left(\frac{\partial\phi}{\partial x'}\right)^2\right\rangle dk'\nonumber\\
&=&\frac{|k'|}{4\pi}dk'.\label{E0k'}
\end{eqnarray}
The total zero-point energy $E_0$ defined by \eqref{totoal zero} in the $(t, x)$ system is transformed to $E'_0$ in the $(t', x')$ system:
\begin{equation}\label{total zero'}
E'_0=\int_{-\infty}^{+\infty}E_0^{'(k', \,k'+dk')}=\int_{-\infty}^{+\infty}\frac{|k'|}{4\pi}dk'.
\end{equation}

The integrand $\frac{|k'|}{4\pi}$ in the above expression \eqref{total zero'} for $E'_0$ in the $(t', x')$ system appears identical to the integrand $\frac{|k|}{4\pi}$ in the expression \eqref{totoal zero} for $E_0$ in the $(t, x)$ system. For this reason, one might be tempted to conclude that $E'_0=E_0$ and the zero-point energy is Lorentz invariant.

However, both integrals in \eqref{totoal zero} and \eqref{total zero'} are divergent. Mathematically, comparing two infinities can be subtle. A famous example that highlights this subtlety is Hilbert's paradox of the Grand Hotel. Thus, caution must be exercised when comparing the divergent quantities $E_0$ and $E'_0$.

In fact, using the relation \eqref{Doppler shift} between $k$ and $k'$ to compare \eqref{E0k} and \eqref{E0k'}, we obtain that $E_0^{'(k', \,k'+dk')}$ observed in the $(t', x')$ system is related to $E_0^{(k, \,k+dk)}$ observed in the $(t, x)$ system by
\begin{equation}\label{boost energy}
E_0^{'(k', \,k'+dk')}=
\begin{cases}
    \left(\frac{1+v}{1-v}\right)E_0^{(k, \,k+dk)},  & \quad \text{if } k<0,\\
    \left(\frac{1-v}{1+v}\right)E_0^{(k, \,k+dk)},  & \quad \text{if } k>0.
  \end{cases}
\end{equation}

The above result \eqref{boost energy} shows that, if $v>0$, for the left (right) moving field modes with $k,\, k'<0$ ($k, \,k'>0$), the zero-point energy $E_0^{'(k', \,k'+dk')}$ observed in the $(t', x')$ system is $\frac{1+v}{1-v}$ ($\frac{1-v}{1+v}$) times larger (smaller) than the corresponding zero-point energy $E_0^{(k, \,k+dk)}$ observed in the $(t, x)$ system. Physically, this result is easy to understand. The right moving observer with $v>0$ would see the left (right) moving field modes to be blue (red) shifted to higher (lower) energy by the factor $\left(\frac{1+v}{1-v}\right)^{\frac{1}{2}}$ ($\left(\frac{1-v}{1+v}\right)^{\frac{1}{2}}$) and at the same time the width of the frequency range is enlarged (narrowed) from $(k, k+dk)$ to $(k', k'+dk')$ by the same factor $\left(\frac{1+v}{1-v}\right)^{\frac{1}{2}}$ ($\left(\frac{1-v}{1+v}\right)^{\frac{1}{2}}$).

Therefore, it is evident that $E_0^{(k, \,k+dk)}\neq E_0^{'(k', \,k'+dk')}$ and the zero-point energy produced by the field modes with wave numbers between $k$ and $k+dk$ is not Lorentz invariant. Then the total zero-point energy $E_0$, which is the integration of $E_0^{(k, \,k+dk)}$ over all $k$ from $-\infty$ to $+\infty$, should also not be Lorentz invariant.

In fact, since the total zero-point energy comes from the zero-point fluctuations of the left moving field modes and the right moving field modes
\begin{eqnarray}
E_0&=&E_0^{\mathrm{L}}+E_0^{\mathrm{R}},\\
E'_0&=&E_0^{'\mathrm{L}}+E_0^{'\mathrm{R}},\label{zero energy split}
\end{eqnarray}
where
\begin{eqnarray}
E_0^{\mathrm{L}}&=&\int_{-\infty}^0E_0^{(k, \,k+dk)},\\
E_0^{\mathrm{R}}&=&\int^{+\infty}_0E_0^{(k, \,k+dk)},\\
E_0^{'\mathrm{L}}&=&\int_{-\infty}^0E_0^{'(k', \,k'+dk')},\\
E_0^{'\mathrm{R}}&=&\int^{+\infty}_0E_0^{'(k', \,k'+dk')},
\end{eqnarray}
and, from \eqref{boost energy} we have
\begin{equation}\label{energy boost relation}
E_0^{'\mathrm{L}}=\left(\frac{1+v}{1-v}\right)E_0^{\mathrm{L}},\quad E_0^{'\mathrm{R}}=\left(\frac{1-v}{1+v}\right)E_0^{\mathrm{R}}.
\end{equation}
Inserting the above relations \eqref{energy boost relation} into \eqref{zero energy split} we obtains that
\begin{eqnarray}\label{Lorentz violation}
E'_0&=&\left(\frac{1+v}{1-v}\right)E_0^{\mathrm{L}}+\left(\frac{1-v}{1+v}\right)E_0^{\mathrm{R}}\nonumber\\
&=&E_0+\left(\frac{2v}{1-v}\right)E_0^{\mathrm{L}}-\left(\frac{2v}{1+v}\right)E_0^{\mathrm{R}}\nonumber\\
&\neq &E_0,
\end{eqnarray}
which implies that the total zero-point energy of the Minkowski vacuum is not Lorentz invariant.

It is interesting to observe that, if one still insists that the zero-point energy is Lorentz invariant, i.e., if one insists that $E'_0=E_0$, then the above relation \eqref{Lorentz violation} leads to $E'_0=E_0\equiv 0$. This gives an explanation for the result of zero contribution from the massless fields to the vacuum energy, which is obtained by evaluating the divergent integral \eqref{vacuum energy} using Lorentz invariant regularization methods such as dimensional regularization. Our analysis presented above suggests that zero-point fluctuation is not Lorentz invariant. It is not surprising that dimensional regularization yields zero zero-point energy for massless fields\footnote{For massive fields, since there is a potential term which contributes to the stress energy in the Lorentz invariant form $-\frac{1}{2}m^2\phi^2\eta_{\mu\nu}$, dimensional regularization captures this contribution and does not give zero result, unlike the case of massless fields.}, since any attempt to regularize \eqref{vacuum energy} in a Lorentz invariant way would miss all the contributions from the non-Lorentz invariant fluctuations.

\subsection{Four Possible Cases}\label{four cases}
Our discussions in the last subsection suggest that $\left\langle\rho^{(\mathrm{vac})}\right\rangle$ might be negative and the vacuum equation of state \eqref{veos} might be violated. This leads to four different possible cases for the gravitational property of the expectation value of the matter fields vacuum stress energy tensor:

\emph{Case I:} \,\,The vacuum equation of state \eqref{veos} is satisfied, and Boson fields contribute more, resulting in $\left\langle T_{00}\right\rangle>0$;

\emph{Case II:} \,\,The vacuum equation of state \eqref{veos} is satisfied, and Fermi fields contribute more, resulting in $\left\langle T_{00}\right\rangle<0$;

\emph{Case III:} \,\,The vacuum equation of state \eqref{veos} is violated, and Boson fields contribute more, resulting in $\left\langle T_{00}\right\rangle>0$;

\emph{Case IV:} \,\,The vacuum equation of state \eqref{veos} is violated, and Fermi fields contribute more, resulting in $\left\langle T_{00}\right\rangle<0$.

Next we investigate the small scale structure of the spacetime for these four cases.

\subsubsection{Small Scale Spacetime Structure for Cases I and II}
\paragraph{Case I.}  Our discussions in the last section \ref{ccp} have assumed that \eqref{veos} is satisfied and $\left\langle T_{00}\right\rangle>0$, corresponding to \emph{case I}. From \eqref{dominance} we have the shear attraction generated by the zero-point fluctuations in the Riemann curvature dominates over the repulsion generated by the zero-point fluctuations in the quantum matter fields. In this scenario, \eqref{evolution of the scale factor oscillator} describes an oscillator with varying frequencies, and the local scale factor $a(t, \mathbf{x})$ oscillates around $0$. Note that $a(t, \mathbf{x})$ crosses $0$ and takes both positive and negative values, but the volume element $\sqrt{h}\,d^3x=|a|^3d^3x$ always takes non-negative values, which means that locally the average size $|a|$ of space at each point exhibits an eternal series of oscillations between expansion and contraction. During each cycle of oscillation, locally the space collapses to zero size and immediately rebounds.

The large shear $\sigma^2$ implies significant local anisotropy in spacetime. Note that the local scale factor $a(t, \mathbf{x})$, defined by $h=a^6$, incorporates contributions from all metric components and represents the local size of space at each spatial point $\mathbf{x}$ after averaging over all directions. A more accurate picture is that space expands in certain directions and contracts in others, with the directions of expansion and contraction constantly changing. An initial small sphere will quickly distort toward an ellipsoid with principle axes given by the eigenvectors of $\sigma^i_j$, with rate given by the eigenvalues of $\sigma^i_j$ (see e.g. Wald 1984, \cite{Wald:1984rg} p.218). More precisely, according to the analysis of Belinski, Khalatnikov, and Lifshitz (BKL) \cite{Belinsky:1982pk, Belinsky:1970ew, Belinskii:1972sg, PhysRevLett.24.76}, locally the spacetime exhibits chaotic, oscillatory ``mixmaster" \cite{PhysRevLett.22.1071, PhysRev.186.1319} behavior. The dynamics is not a simple monotonic collapse in most cases --- instead, directions of space take turns collapsing or expanding in rapid succession.

The structure of the spacetime at relatively larger scales would be these small scale structures ``glued" together. Due to the random quantum fluctuations, the phases of the oscillations of $a(t, \mathbf{x})$ vary from point to point. As we discussed in the last section \ref{ccp}, when averaging over macroscopic scales, these wild oscillations in $a(t, \mathbf{x})$ and the corresponding local expansions and contractions of space almost cancel each other out. However, due to a weak parametric resonance effect produced by the zero-point fluctuations, expansions slightly outweigh contractions. This tiny net expansion accumulates on cosmological scale, ultimately leading to the observed slowly accelerating expansion of the universe. One issue about this picture is the formation of singularities at $a(t,\mathbf{x})=0$. This issue has been extensively discussed in Sec. IX B \cite{PhysRevD.95.103504} and Sec. VIII C of \cite{PhysRevD.102.023537} of our previous papers.

\paragraph{case II.} In this case, the vacuum equation of state \eqref{veos} is satisfied, leading to
\begin{equation}\label{sign with satisfication}
\left\langle\Omega_{\mathrm{mat}}^2\right\rangle=-\frac{8\pi G}{3}\left\langle T_{00}\right\rangle=-w^2>0.
\end{equation}
The local conservation of energy and momentum \eqref{conservation equation} requires that $\left\langle\Omega_{\mathrm{mat}}^2\right\rangle\equiv \mathrm{constant}$ all over the spacetime.

So the zero-point fluctuations in the matter fields produce gravitational attractions. In this case, \eqref{dominance} remains valid and the evolution of $a(t,\mathbf{x})$ and the small scale structure of the spacetime resemble \emph{case I}. The only difference is that the oscillation of $a(t,\mathbf{x})$ occurs at a slightly faster rate, as both the zero-point fluctuations in the matter fields and in the gravitational field contribute attractive effects. As a result, the oscillation frequency $\sqrt{\Omega_{\mathrm{mat}}^2+2\sigma^2/3}$ of $a(t,\mathbf{x})$ is slightly larger in this case.

\subsubsection{Small Scale Spacetime Structure for Cases III and IV}
If the vacuum equation of state \eqref{veos} is violated, then according to the expressions \eqref{vacuum energy} and \eqref{vacuum pressure}, $\langle T_{00}\rangle$ and $\langle T_{ii}\rangle\,(i=1, 2, 3)$ should have the same sign. The crucial distinction between \emph{cases III} and \emph{IV} compared to \emph{cases I} and \emph{II}, where \eqref{veos} is satisfied, is that the expectation values $\langle T_{00}\rangle$, $\langle T_{ii}\rangle\,(i=1, 2, 3)$ and $\left\langle\Omega_{\mathrm{mat}}^2\right\rangle$ are not constant\footnote{Note that the constant expressions \eqref{vacuum energy} and \eqref{vacuum pressure} for $\langle T_{00}\rangle$, $\langle T_{ii}\rangle\,(i=1, 2, 3)$ are calculated in the fixed Minkowski spacetime. In principle, one should perform the calculations in the dynamically evolving spacetime and the values of $\langle T_{00}\rangle$ and $\langle T_{ii}\rangle$ would depend on the local scale factor $a(t, \mathbf{x})$ (and also all other metric components).}. The local conservation of energy and momentum \eqref{conservation equation} requires that they must depend on the local scale factor $a(t, \mathbf{x})$ and all other metric components. 

For \emph{cases I} and \emph{II}, since $\left\langle\Omega_{\mathrm{mat}}^2\right\rangle$ is constant, we always have
$\Omega_{\mathrm{mat}}^2+2\sigma^2/3>0$ and thus the local scale factor $a(t, \mathbf{x})$, whose evolution is described by \eqref{evolution of the scale factor oscillator}, must cross $0$ and take both positive and negative values. However, for \emph{cases III} and \emph{IV}, since $\Omega_{\mathrm{mat}}^2$ varies as the spactime evolves, $\Omega_{\mathrm{mat}}^2+2\sigma^2/3$ can become negative.

More specifically, as $|a|$ decreases (increases), $\langle T_{00}\rangle$ and $\langle T_{ii}\rangle$ increase\footnote{Imagine a box of gas of volume $|a|^3$ which satisfies the equation of state $p=\frac{1}{3}\rho>0$, as $|a|$ decreases, the environment would do positive (negative) work to the gas so that we have the energy density $\rho$ increases (decreases).} (decrease) so that the gravitational attraction from $\left\langle\Omega_{\mathrm{mat}}^2\right\rangle>0$ becomes stronger (weaker) as the space locally contracts (expands).

In this scenario, we also have that the factor $\Omega_{\mathrm{mat}}^2+2\sigma^2/3$ on the right hand side of evolution equation \eqref{evolution of the scale factor oscillator} for $a(t, \mathbf{x})$ is positive. As a result, the evolution of $a(t,\mathbf{x})$ and the small scale structure of the spacetime also resemble \emph{cases I} and \emph{II}, except the differences in the exact oscillation details of $a(t,\mathbf{x})$ caused by the varying $\left\langle\Omega_{\mathrm{mat}}^2\right\rangle$.

\subsubsection{Small Scale Spacetime Structure for the Case IV}
In \emph{case IV}, we have $\langle T_{00}\rangle<0$, $\langle T_{ii}\rangle<0,\,(i=1, 2, 3)$ and $\left\langle\Omega_{\mathrm{mat}}^2\right\rangle<0$. Consequently, the zero-point fluctuations in the matter fields generate gravitational repulsion. 

In this scenario, as $|a|$ decreases (increases), $\langle T_{00}\rangle$ and $\langle T_{ii}\rangle$ would decrease\footnote{Imagine a box of gas of volume $|a|^3$ which satisfies the equation of state $p=\frac{1}{3}\rho<0$, as $|a|$ decreases (increases), the environment would do negative (positive) work to the gas so that we have the energy density $\rho$ decreases (increases).} (increase). As a result, the gravitational repulsion from $\left\langle\Omega_{\mathrm{mat}}^2\right\rangle<0$ becomes stronger (weaker) as the space locally contracts (expands). Moreover, we would have
\begin{equation}\label{case iv property}
\left\langle\Omega_{\mathrm{mat}}^2\right\rangle\to-\infty, \quad\text{as}\quad |a|\to 0,
\end{equation}
meaning that the gravitational repulsion generated by the matter fields becomes infinitely large if locally the space shrinks to zero size.

For \emph{cases I, II} and \emph{III}, we always have
\begin{equation}
\Omega_{\mathrm{mat}}^2+\frac{2\sigma^2}{3}>0,
\end{equation}
and thus the local scale factor $a(t, \mathbf{x})$, whose evolution is described by \eqref{evolution of the scale factor oscillator}, must cross $0$ and take both positive and negative values. However, for \emph{case IV}, due to the property \eqref{case iv property} of $\left\langle\Omega_{\mathrm{mat}}^2\right\rangle$, the situation might become different as $|a(t, \mathbf{x})|$ becomes sufficiently small. As $a(t, \mathbf{x})$ approaches zero, the absolute value of $\left\langle\Omega_{\mathrm{mat}}^2\right\rangle$ becomes arbitrarily large. So it is possible that the repulsion from $\left\langle\Omega_{\mathrm{mat}}^2\right\rangle$ becomes larger than the shear attraction generated by the zero-point fluctuations in the Riemann curvature, leading to
\begin{equation}\label{near singularity behavior}
\Omega_{\mathrm{mat}}^2+\frac{2\sigma^2}{3}<0,\quad\text{as}\quad |a|\to 0.
\end{equation}

Whether the above inequality \eqref{near singularity behavior} is valid needs more investigations, since the behavior of $\sigma^2$ as $|a|$ approaches $0$ is not clear so far. The evolution of $\sigma^2$ is given by (see Eq. (84) in our previous paper \cite{PhysRevD.102.023537}) 
\begin{equation}
(\sigma^2)^{\cdot}=-\frac{6\dot{a}}{a}\sigma^2-R^{(3)}_{ij}\sigma^{ij}+8\pi GT_{ij}\sigma^{ij}.
\end{equation}
It seems that $\sigma^2$ also goes to infinity as $|a|$ approaches $0$. If $\sigma^2$ grows faster than $|\left\langle\Omega_{\mathrm{mat}}^2\right\rangle|$ as $|a|\to 0$, the inequality \eqref{near singularity behavior} may not hold. 

If, somehow, the inequality \eqref{near singularity behavior} is valid, the repulsion from $\left\langle\Omega_{\mathrm{mat}}^2\right\rangle$ could potentially prevent $a(t, \mathbf{x})$ from collapsing to the singularities at $a(t, \mathbf{x})=0$. Instead, for each cycle, it would collapse to some minimum value\footnote{Note that due to the randomness of quantum fluctuations, the minimus value $a_{\mathrm{min}}$ may take different values for different cycles of oscillations.} at $a(t, \mathbf{x})=a_{\mathrm{min}}>0$ and then bounce back\footnote{Further investigation confirms that the negative fermionic zero-point stress-energy generates sufficient gravitational repulsion to prevent the formation of spacetime singularities. A detailed account of these results will be published in due course.}. In this case, $a(t, \mathbf{x})$ would not go across $0$ to take negative values like in the other three cases; its value could always be positive.

The small scale structure of the spacetime in this scenario is also similar to the \emph{cases I, II} and \emph{III}, with the distinction that if \eqref{near singularity behavior} holds, the singular bounces at the singularities $a(t, \mathbf{x})=0$ in \emph{cases I, II} and \emph{III} are replaced by smooth bounces at $a(t, \mathbf{x})=a_{\mathrm{min}}>0$ in the \emph{case IV}.

To summary, the four possible cases exhibit similar small scale spacetime structures, and our proposal works for all of them.

\subsection{What if $\Lambda_{\mathrm{mat}}>E_p$?}
While it is widely anticipated that the quantum field theory description of matter breaks down at energy scales beyond the Planck scale, where a cutoff $\Lambda_{\mathrm{mat}}\leq E_p$ is expected, there is currently no satisfactory quantum theory of gravity to prove this expectation, nor are there experiments capable of reaching Planck scale to verify it. So what if quantum field theory is still valid beyond Planck scale and we have $\Lambda_{\mathrm{mat}}>E_p$?

In such a scenario, the inequality \eqref{inequality} is no longer valid. However, we have
\begin{equation}
G^2\Lambda_{\mathrm{mat}}^6>G\Lambda_{\mathrm{mat}}^4,
\end{equation}
which means that the contribution to the extra attraction from the zero-point fluctuation of the matter fields vacuum surpasses the repulsion generated from the large zero-point energy of the matter field vacuum itself. Thus, in this situation, the catastrophic explosion of the universe could also be prevented.

\subsection{Differences from Previous Works}\label{differences}
One crucial distinction between this paper and our earlier papers \cite{PhysRevD.95.103504}, \cite{PhysRevD.98.063506}, \cite{PhysRevD.102.023537} and \cite{PhysRevLett.125.051301} on the cosmological constant problem is that we did not consider the extra attraction generated automatically by the quantum zero-point fluctuations in Riemann curvature, which originates from the nonlinear nature of gravity. In the papers \cite{PhysRevD.95.103504} and \cite{PhysRevD.98.063506}, the gravitational effect of the matter fields we considered are already attractive, while in the papers \cite{PhysRevD.102.023537} and \cite{PhysRevLett.125.051301}, we added a large negative bare cosmological constant in the Einstein field equations to provide the needed attraction.

More concretely, our initial paper \cite{PhysRevD.95.103504} considered a simplified inhomogeneous but locally isotropic metric $ds^2=-dt^2+a^2(t, \mathbf{x})\left(dx^2+dy^2+dz^2\right)$, with the evolution of $a(t, \mathbf{x})$ given by
\begin{equation}\label{initial system}
\ddot{a}+\Omega^2_{\mathrm{mat}}a=0.
\end{equation}
The source of gravity is a massless scalar field $\phi$ with $\Omega^2_{\mathrm{mat}}\propto T_{00}+\frac{1}{a^2}\left(T_{11}+T_{22}+T_{33}\right)=2\dot{\phi}^2>0$, which yields gravitational attraction. In this locally isotropic metric, the shear term is zero, thus the extra shear attraction studied in this paper is not included. The later work presented in \cite{PhysRevD.98.063506} considered the same simplified metric but with multiple scalar fields as the matter source of gravity. The matter fields vacuum \cite{PhysRevD.95.103504} and \cite{PhysRevD.98.063506} has positive energy density and violated the vacuum equation of state. Therefore, the matter fields we used in \cite{PhysRevD.95.103504} and \cite{PhysRevD.98.063506} correspond to the \emph{case III} we discussed in the last subsection \ref{four cases}.

Our recent work presented in \cite{PhysRevD.102.023537} and \cite{PhysRevLett.125.051301} also adopted the same inhomogeneous and locally anisotropic metric given by \eqref{general metric}. The key difference is that in \cite{PhysRevD.102.023537} and \cite{PhysRevLett.125.051301}, we added a bare cosmological constant $\lambda_B$ in the Einstein field equations and the evolution equation for $a(t, \mathbf{x})$ becomes
\begin{equation}
\ddot{a}+\left(-\frac{1}{3}\lambda_B+\Omega_{\mathrm{mat}}^2+\frac{2\sigma^2}{3}\right)a=0.
\end{equation}
Here, the source of gravity is supposed to include all (known and unknown) fundamental fields in nature, with the sign of $\langle\Omega_{\mathrm{mat}}^2\rangle$ undetermined. In this scenario, the shear term $\sigma^2>0$ is not zero; however, we did not realize that the quantum zero-point fluctuations in the Riemann curvature could naturally generate a significant $\sigma^2$. In order to achieve the necessary strong attraction to prevent the catastrophic explosion if $\langle\Omega_{\mathrm{mat}}^2\rangle<0$, we took the bare cosmological constant $\lambda_B$ to exceedingly large negative values such that $-\lambda_{B}\gg -\langle\Omega_{\mathrm{mat}}^2\rangle$.

Lozano and Mazzitelli \cite{Lozano:2020xga} investigated the role of noise produced by the fluctuations of the energy momentum tensor of the matter fields vacuum around its mean value in the evolution of the local scale factor introduced in our initial work \cite{PhysRevD.95.103504}. They utilized a simplified toy model, wherein the equation of motion for the scale factor took the form (Eq.(38) in \cite{Lozano:2020xga})
\begin{equation}\label{Lozano equation}
\ddot{a}-\frac{8\pi G}{3}\left(\left\langle\rho\right\rangle+\xi_{\rho}\right)a=0.
\end{equation}

They proposed that the noise $\xi_{\rho}$ in equation \eqref{Lozano equation} has an effect similar to the dynamic stabilization in Kapitza's pendulum \cite{Kapitza}. There is some similarity between equation \eqref{Lozano equation} and the geodesic deviation equation \eqref{matrix form}, with the curvature matrix $\mathrm{R}$ given by \eqref{2}. The mechanism in this paper that produces the Ponderomotive-like tidal forces between neighboring geodesics has similar physical origin to the mechanism stabilizing Kapitza's pendulum.

Besides the differences in the technical details, one of the key distinctions between the work presented in this paper and in \cite{Lozano:2020xga} is that, in this paper the active zero-point fluctuations of the gravitational vacuum in the Weyl curvature is crucial, while in \cite{Lozano:2020xga} only the fluctuations of the passive zero-point fluctuations produced by the matter fields vacuum is considered. Specifically, their analysis did not include the zero-point fluctuation induced large shear $\sigma^2$, which plays a critical role in our study.

\section{Summary}\label{summary}
By analyzing the dynamics of the Jacobi field described by the geodesic deviation equation \eqref{matrix form} in the free falling observer $\gamma$'s local inertial frame, we have presented a general result indicating that Riemann curvature fluctuations give rise to extra ponderomotive-like tidal attraction between neighboring geodesics. By investigating the spacetime described by the locally inhomogeneous and anisotropic metric \eqref{general metric}, we have established that Riemann curvature fluctuations inevitably induce nonzero shear attraction, which tends to shrink the local scale factor $a(t, \mathbf{x})$ whose evolution is given by \eqref{evo}. These findings imply that the nonlinear dynamics of curvature fluctuations inherently bend the background geometry in a manner that generates extra attraction.

If one is to believe the quantum principle and Einstein's theory, the spacetime geometry everywhere experiences unavoidable, natural, zero-point fluctuations. These zero-point fluctuations naturally give rise to the aforementioned additional tidal and shear attraction. The conventional formulation of the cosmological constant problem solely considers the gravitational repulsion produced by the matter fields vacuum with positive energy and negative pressure, but overlooks the attraction generated by these zero-point fluctuations in the spacetime geometry. However, it turns out that this extra attraction plays a crucial role. The enormous gravitational repulsion produced by the matter fields vacuum is completely suppressed by this more substantial attraction. As a result, the predicted catastrophic explosion of the universe caused by the huge positive energy and negative pressure of the matter fields vacuum is avoided. Furthermore, the zero-point fluctuations of the spacetime itself could drive the observed slow acceleration of the universe's expansion through a subtle parametric resonance effect.

This research demonstrates that the substantial zero-point energy of the quantum matter fields vacuum does not necessarily result in catastrophic consequences, as long as the substantial zero-point fluctuations of the gravitational field are also taken into account. The zero-point energy could be genuinely as large as predicted by quantum field theory, and it does gravitate in accordance with the equivalence principle. This result suggests that the substantial zero-point fluctuations of both the matter fields and the gravitational field should play a significant role in the construction of a future successful quantum theory of gravity --- a consideration often overlooked in existing attempts to quantize gravity.

\section*{Acknowledgements}
I would like to specially thank William G. Unruh for extensive discussions and detailed suggestions on this research and earlier versions of this paper. I would also like to thank Philip C.E. Stamp, Zhen Zhu and Shao-jiang Wang for valuable discussions. This research is supported by the ``Young Talent Support Plan" of Xi'an Jiaotong University (Grant Nos. 012400/71211201010703, 012400/71211222010707, 012400/11304222010712, 012400/11301223010705) and the ``Young Talent Introduction Plan" of Shaanxi Province (Grant No. 050700/71240000000097).

\bibliographystyle{unsrt}
\bibliography{References}

\end{document}